
\input phyzzx
\singlespace

%
\FIG\QCP{
The static quark potential $V(r)$
in the dual Ginzburg-Landau theory
in the quenched approximation.
The horizontal axis denotes the distance between a static
quark-antiquark pair.
The solid curve is the result for e=5.5, $m_B$=500 MeV and
$m_\chi $=1.26 GeV.
The dashed curve denotes the Cornell potential
with $e_{_C}=2.0$ and $k_{_C}=1.0$GeV/fm.
}
%
\FIG\PCR{
The normalized pair creation rate $w(p_{_T})/w_{sc}$
as a function of the energy of the created $q$-$\bar q$
pair for $M$=350 MeV and $k$=1.0GeV/fm.
The horizontal axis denotes the quark pair energy,
$2E_q(p_{_T})=2(p_{_T}^2+M^2)^{1/2}$.
The expectation value of the quark pair energy is
found to be $\langle 2E_q(p_{_T}) \rangle  \simeq 850{\rm MeV}$.
}
%
\FIG\SCE{
The infrared screening effect on the quark confining potential
$V_{\rm linear}^{sc}(r)$
for several values of the infrared cutoff $\epsilon $.
The QCD-monopole mass $m_\chi $ is chosen to
reproduce the string tension, $k$=1.0GeV/fm :
$m_\chi =$1.26, 1.15 and 1.01 GeV for $\epsilon $=0, 100 and 200MeV,
respectively.
One finds the saturation behavior of the confining potential
in the long distance.
}
%
%
\FIG\DQM{
The squared value of the dynamical quark mass, $M(p^2)$, as a
function of the Euclidean momentum squared $p^2=-p_{_M}^2$
for $e=5.5$, $\Lambda _{\rm QCD}=200{\rm MeV}$,
$m_B=500{\rm MeV}$ and $\epsilon $=80MeV.
The solid curve in the space-like region, $p^2>0$, denotes the
numerical solution of the Schwinger-Dyson equation.
The solid curve in the time-like region, $p^2<0$, is obtained by
its smooth extrapolation.
The dotted straight line denotes the on-shell state $M^2(p^2)=-p^2$.
One finds the solid curve does not cross this dotted straight line,
which means the disappearance of the pole in the quark propagator.
}
%
\FIG\QEM{
The dynamical quark mass $M(p^2)$ as a
function of the Euclidean momentum squared $p^2=-p_{_M}^2$
for several values of the mass of the dual-gauge field $\vec B_\mu $,
$m_B$=300, 400 and 500 MeV.
The other parameters are set to the same values as in Fig.4.
No nontrivial solution is found for the small values
of $m_B$, e.g.  $m_B \lsim 200 {\rm MeV}$.
The quark mass $M(p^2)$ takes larger value at each $p^2$ as
$m_B$ gets larger, which physically means that
QCD-monopole condensation contributes to the
dynamical chiral-symmetry breaking.
}
%
\FIG\PRM{
The physical quantities related to the dynamical chiral-symmetry
breaking as a function of the gauge coupling constant of QCD, $e$,
for several values of the dual-gauge field $m_B$:
(a) the dynamical quark mass at $p=0$, $M(0)$,
(b) the quark condensate $\langle \bar qq \rangle_{_{\rm RGI}}$, and
(c) the pion decay constant $f_\pi $.
One finds that these quantities are well reproduced for
the parameter set
($e=5.5, \Lambda _{\rm QCD}=200{\rm MeV}$, $m_B=500{\rm MeV}$, $\epsilon
$=80MeV),
which is consistent with the confining property and the
perturbative QCD.
On the contrary, large values for $e$ and $\Lambda _{\rm QCD}$ are needed
to reproduce the magnitude of the chiral-symmetry
breaking for $m_B$=0.
}


\title{ Color Confinement, Quark Pair Creation
and Dynamical Chiral-Symmetry Breaking
in the Dual Ginzburg-Landau Theory }
\vskip 15pt
\centerline{H.~Suganuma$^{\rm a)}$,
S.~Sasaki$^{\rm b)}$ and H.~Toki$^{\rm a, b)}$
}
\vskip 10pt
\centerline{a) {\it The Institute of Physical and Chemical
Research (RIKEN),}}
\centerline{\it Hirosawa 2-1, Wako, Saitama 351-01, Japan}
\centerline{b) {\it Department of Physics,
Tokyo Metropolitan University,}}
\centerline{\it Minami-Ohsawa 1-1, Hachioji, Tokyo 192-03, Japan}
\abstract{
We study the color confinement, the $q$-$\bar q$ pair creation
and the dynamical chiral-symmetry breaking
of nonperturbative QCD by using the dual Ginzburg-Landau theory,
where QCD-monopole condensation plays an essential
role on the nonperturbative dynamics in the infrared region.
As a result of the dual Meissner effect,
the linear static quark potential, which characterizes
the quark confinement, is obtained in the long distance
within the quenched approximation.
We obtain a simple expression for the string tension similar to
the energy per unit length of a vortex in the superconductivity
physics. The dynamical effect of light quarks on the quark confining
potential is investigated in terms of the infrared screening effect
due to the $q$-$\bar q$ pair creation or the cut of the hadronic string.
The screening length of the potential is estimated by using
the Schwinger formula for the $q$-$\bar q$ pair creation.
We introduce the corresponding infrared cutoff to the strong
long-range correlation factor in the gluon propagator
as a dynamical effect of light quarks,
and obtain a compact formula of the quark potential
including the screening effect in the infrared region.
We investigate the dynamical chiral-symmetry breaking
by using the Schwinger-Dyson equation, where the gluon
propagator includes the nonperturbative effect related to
the color confinement.
We find a large enhancement of the chiral-symmetry
breaking when the dual Meissner effect takes place, which supports
the close relation between the color confinement and the
chiral-symmetry breaking.
The dynamical quark mass, the pion decay
constant and the quark condensate are well reproduced by using
the consistent values of the gauge coupling constant and the
QCD scale parameter with the perturbative QCD and the
quark confining potential.
The confinement of light quarks is also investigated
by the smooth extrapolation of the quark mass function
to the time-like momentum region.
We find the disappearance of physical poles in the
light-quark propagator, and show the confinement of light quarks
in the dual Ginzburg-Landau theory.
}

\chapter{Introduction}

The quantum chromodynamics (QCD) has been accepted as the fundamental
theory of strong interactions. Due to the phenomena of the
asymptotic freedom in QCD, the ordinary perturbative technique is
valid and useful for the analysis of the high-energy hadron reactions
with much success
\REF\nachtmann{
For instance, O.~Nachtmann, ``Elementary Particle Physics",
(Springer-Verlag, Berlin, 1990) 1.
}[\nachtmann].
On the contrary, QCD becomes the strong-coupling
gauge theory, and exhibits rich phenomena in relation with
the nonperturbative nature in the low-energy region.
In particular, the color confinement and the
dynamical chiral-symmetry breaking are
unique features in the nonperturbative region of QCD, and
have been studied with much interest by using various strategies,
e.g. the effective-model approach or the lattice QCD.

In the perturbative QCD,
the QCD scale parameter $\Lambda _{\rm QCD}$ is one of the most important
quantities, and provides a typical scale of the strong interaction.
Using the one-loop level perturbation, one obtains
the running coupling constant in QCD at a large momentum scale $p^2$,
$$
\alpha _s (p^2) \equiv {e^2(p^2) \over 4\pi }=
{12\pi  \over (11N_c-2N_f) \ln(p^2/\Lambda ^2_{\rm QCD})},
\eqn\RCCa
$$
where $N_c$ and $N_f$ are the numbers of the color and flavor,
respectively [\nachtmann].
In the infrared region, this running coupling constant becomes large
and formally diverges at $p^2 = \Lambda _{\rm QCD}^2$, so that
the simple perturbation theory is no more meaningful
and some nonperturbative effect should appear instead.
Some experimental data suggests
$\alpha _s(m_Z^2) \simeq 0.108\pm 0.005 $
from the analysis of the recent LEP data
\REF\amaldi{
U.~Amaldi, W.~de~Boer and H.~Furstenau,
Phys.~Lett.~{\bf B260} (1991) 447.
}
[\amaldi], or $\alpha _s((35{\rm GeV})^2)=0.10 \sim 0.17 $
from the high-energy lepton-hadron reaction [\nachtmann].
The QCD scale parameter $\Lambda _{\rm QCD }$ is obtained from
these experimental data, and its recent value is given as
$\Lambda _{\rm QCD}(\overline{\rm MS})=220 \pm 15 \pm 50{\rm MeV}$
in the $\overline{\rm MS}$ scheme [\nachtmann]
from the experimental data of the BCDMS collaboration
\REF\bcdms{
BCDMS Collaboration, A.~C.~Benvenuti et al.,
Phys.~Lett.~{\bf B223} (1989) 490.
}[\bcdms].
Hence, one roughly estimates from Eq.{\RCCa},
$\alpha _s((0.5{\rm GeV})^2)\sim 1$, which means the breakdown of the
validity of the perturbation theory.
Thus, we expect the appearance of the nonperturbative effect
in the infrared region, $p \lsim 1 {\rm GeV}$, due to the strong
coupling.

Now, we shall give an overview of the dynamical chiral-symmetry
breaking and the color confinement as the outstanding nonperturbative
features in the infrared region of QCD.
Although QCD Lagrangian has the chiral symmetry in the massless quark
limit, it is spontaneously broken in the nonperturbative QCD vacuum
\REF\cheng{
For instance,
T.~P.~Cheng and L.~F.~Li,
``Gauge theory of elementary particle physics",
(Clarendon press, Oxford, 1984) 1.
}
[\cheng].
In the phenomenological point of view, the dynamical chiral-symmetry
breaking leads the absence of the parity doubling for hadrons
and the success of the low-energy theorem or the current
algebra
\REF\bando{
For a review article, \nextline
M.~Bando, T.~Kugo and K.~Yamawaki,
Phys.~Rep.~{\bf 164} (1988) 217.
}
[\bando].
The dynamical chiral-symmetry breaking is
characterized by the quark condensate,
$\langle \bar qq \rangle \simeq -(225 \pm 25 {\rm MeV})^3$
\REF\gasser{
J.~Gasser and H.~Leutwyler, Phys.~Rep.~{\bf 87} (1982) 77.
}
\REF\shuryak{
For a review article,
E.~V.~Shuryak, ``The QCD vacuum, hadrons and
the superdense matter", (World Scientific, Singapore, 1988) 1.
}
[\gasser,\shuryak].
Quark condensation is caused by the attractive interaction between
quarks, and this phenomenon is analogous with  Cooper-pair
condensation in the superconductivity.
Such a physical process has been demonstrated by using the effective
models of QCD, for instance, the Nambu-Jona-Lasinio model
\REF\nambuA{
Y.~Nambu and G.~Jona-Lasinio, Phys.~Rev.~{\bf 122} (1961) 345.
}[\nambuA]
or the instanton liquid model [\shuryak].
The SU(6) theory [\nachtmann] or the effective-model
approaches [\shuryak] suggest that light quarks get
a large effective mass, $M \simeq 350{\rm MeV}$,
and behave as massive constituent quarks
in the infrared energy region, as the result of the dynamical
chiral-symmetry breaking.
The pion is identified as the Nambu-Goldstone boson
related to the dynamical chiral-symmetry breaking,
and obeys the low-energy theorem or the current algebra [\bando],
where the pion decay constant, $f_\pi  \simeq 93 {\rm MeV}$, is
also a relevant quantity characterizing the breaking of
chiral symmetry [\cheng].
Thus, the dynamical chiral-symmetry breaking is characterized by
several quantities, the chiral condensate
$\langle \bar qq \rangle \simeq -(225 \pm 25{\rm MeV})^3$,
the effective quark mass $M \simeq 350{\rm MeV}$, and
the pion decay constant $f_\pi \simeq 93{\rm MeV}$.

The color confinement is signaled as the absence of the asymptotic
states of the colored particles.
In the phenomenological point of view,
hadrons except the pion are considered to be string-like because
of the Regge trajectories of hadrons
\REF\sailer{For a recent review article,
K.~Sailer, Th.~Sch\"onfeld, Zs.~Schram, A.~Sch\"afer and W.~Greiner,
 J.~Phys.~G: Nucl.~Part.~Phys.~{\bf 17} (1991) 1005.} [\sailer]
or the duality of the hadron reaction.
The string picture of hadrons suggests a linear potential
between quarks, due to which quarks are confined.
In particular, the universality of the Regge slopes of hadrons
means the universal value of the string tension [\sailer],
$k \simeq 1 {\rm GeV/fm}$, which characterizes the strength of
the confinement.
Then, the question on the confinement is how and why the
color-electric flux between quarks is squeezed like a string or a tube.
The natural explanation is the exclusion of the color-electric field
in the nonperturbative QCD vacuum, where
the color dielectric constant vanishes.

In terms of the vanishing of the color dielectric constant,
Y.~Nambu, 't~Hooft and Mandelstam paid attention to an analogy
between the color confinement and the Meissner effect,
the vanishing of the magnetic permeability and the exclusion of
the magnetic flux in the superconductor,
and regarded the color confinement as
the dual version of the superconductivity
\REF\nambuB{
Y.~Nambu, Phys.~Rev.~{\bf D10} (1974) 4262.
}
\REF\thooftA{
G.~'t~Hooft, ``High Energy Physics", ed. A.~Zichichi
(Editorice Compositori, Bologna, 1975).
}
\REF\mandelstam{
S.~Mandelstam, Phys.~Rep.~{\bf C23} (1976) 245.
}
[\nambuB-\mandelstam].
Their idea is based on the duality, a kind of the symmetric property
between the electric part and the magnetic part
in the gauge theories like QED or QCD.
In this picture, the dual Meissner effect,
the exclusion of the (color-)electric field, is brought by
(color-)monopole condensation instead of
Cooper-pair condensation in the superconductivity.

The appearance of magnetic monopoles
in QCD was investigated by 't~Hooft based on the abelian gauge fixing
\REF\thooftB{
G.~'t~Hooft, Nucl.~Phys.~{\bf B190} (1981) 455.
}[\thooftB],
which was assumed to be the relevant gauge fixing to understand the
color confinement.
The abelian gauge fixing is defined by the diagonalization of an
arbitrary gauge-dependent physical variable,
and nonabelian gauge theories are generally reduced into
abelian gauge theories with magnetic monopoles in this gauge.
When such QCD-monopoles are condensed due to some attractive force
between them, the dual Meissner effect takes place, that is, the
color-electric field is excluded in the QCD vacuum.
In this case, the color-electric flux between static quarks are
squeezed into an one-dimensional string or a tube like a vortex
in the superconductivity.
Such a string picture of hadrons leads to a linear potential
between static quarks, characterizing the quark confinement,
and is consistent with the Regge trajectories of hadrons and the
lattice QCD data of the quenched approximation.
Thus, QCD-monopole condensation seems to give a physical
explanation of the color confinement and related phenomena
in the nonperturbative region of QCD.

The Kanazawa group recently formulated the dual Ginzburg-Landau
theory as a phenomenological theory for the nonperturbative QCD
by introducing the QCD-monopole field and its interaction
\REF\suzuki{
T.~Suzuki, Prog.~Theor.~Phys. {\bf 80} (1988) 929 ;
{\bf 81} (1989) 752.
}
\REF\maedanA{
S.~Maedan and T.~Suzuki, Prog.~Theor.~Phys. {\bf 81} (1989) 229.
}
[\suzuki,\maedanA]
based on 't~Hooft's conjecture of QCD-monopole condensation.
Within the quenched approximation,
they derived the linear quark potential in the long distance,
and demonstrated the realization of the dual Meissner effect,
e.g., formation of the hadron flux tube between static quarks
\REF\maedanB{
S.~Maedan, Y.~Matsubara and T.~Suzuki,
Prog.~Theor.~Phys.~{\rm 84} (1990) 130. \nextline
S.~Kamizawa, Y.~Matsubara, H.~Shiba and T.~Suzuki,
Nucl.~Phys.~{\rm B389} (1993) 563.
}
[\maedanB].

In recent years, several authors have pointed out
the {\it abelian dominance}, {\it i.e.}
the importance of the abelian degrees of freedom for the
nonperturbative quantities in the nonabelian
gauge theory based on the lattice gauge simulation
\REF\yotsuyanagi{
T.~Suzuki and I.~Yotsuyanagi,
Phys.~Rev.~{\bf D42} (1990) 4257. \nextline
T.~Suzuki and I.~Yotsuyanagi,
Nucl.~Phys.~{\bf B} (Proc.~Suppl.) {\bf 20} (1991) 236.}
\REF\hioki{S.~Hioki, S.~Kitahara, S.~Kiura, Y.~Matsubara,
O.~Miyamura, S.~Ohno and T.~Suzuki,
Phys.~Lett.~{\bf B272} (1991) 326. \nextline
S.~Hioki, S.~Kitahara, S.~Ohno, T.~Suzuki, Y.~Matsubara
and O.~Miyamura, Phys.~Lett.~{\bf B285} (1992) 343.
}
[\yotsuyanagi, \hioki].
In their studies, the abelian configurations seem to play a
dominant role in the nonperturbative quantities like the Wilson loop
or the Polyakov loop [\yotsuyanagi,\hioki],
and such an evidence of the abelian dominance
seems to lead the relevance of the abelian gauge
to the color confinement.

The important role of the QCD-monopole to the confinement
have been also studied by using the lattice gauge simulations
[\hioki].
In the compact QED, there is the confining phase in the
strong-coupling region
\REF\daggoto{
E.~Daggoto and J.~Kogut, Nucl.~Phys.~{\bf B295}
(1988) 123 and references therein.
}
[\daggoto]
as well as QCD,
and monopole condensation is known to play an essential role
to the confinement from the studies of the lattice gauge theory
\REF\banks{
T.~Banks, R.~Myerson and J.~Kogut,
Nucl.~Phys.~{\bf B129} (1977) 493. \nextline
J.~L.~Cardy, Nucl.~Phys.~{\bf B170} (1980) 369.
}
\REF\degrand{
T.~A.~DeGrand and D.~Toussaint, Phys.~Rev.~{\bf D22} (1980) 2478.
}[\banks, \degrand].
In the nonabelian gauge theories,
several authors have pointed out the importance
of the QCD-monopole to the confinement.
Kronfeld et al.
\REF\kronfeld{
A.~S.~Kronfeld et al., Phys.~Lett.~{\bf B198} (1987) 516. \nextline
A.~S.~Kronfeld, G.~Schierholz and U.~-J.~Weise,
Nucl.~Phys.~{\bf B293} (1987) 461.
}
[\kronfeld]
compared the SU(2) gauge theory and the compact QED
by using the lattice gauge simulation,
and showed that monopole condensation seems
to take place also in the SU(2) gauge theory.
In Ref.[\hioki], an enhancement of the time component of the
monopole current in the deconfinement phase
was reported as an evidence of QCD-monopole condensation,
which suggests the interesting correspondence between the
spatial correlation of the monopoles and the color confinement.
Very recently, the Kanazawa group showed
that the entropy of an extended monopole loop dominates
over its energy and monopole condensation occurs
on the renormalized lattice in the infrared region, $\beta <\beta _c$,
by using the SU(2) lattice gauge simulation
\REF\shiba{
H.~Shiba and T.~Suzuki, Kanazawa Univ., Report No.Kanazawa 93-9, (1993);
Report No.Kanazawa 93-10, (1993).
}[\shiba].
They also showed that the string tension of the SU(2) gauge theory
are well reproduced by extended monopole contributions alone
[\shiba].
Thus, several evidences on the abelian dominance and the
relevant role of the QCD-monopole to the confinement
have been found in the lattice gauge simulation.
Of course, much more progress should be made to understand the
relation between QCD-monopole condensation and the color
confinement.

The dynamical effect of light quarks is also important for the color
confinement, because the linear quark potential is screened
due to the $q$-$\bar q$ pair creation in the long distance,
which means the cut of hadronic strings.
The static quark potential seems to be saturated
in the infrared region, where the system is described as
two mesons including light and heavy fermions.
Such a tendency has been observed in the lattice QCD simulation
with dynamical quarks
\REF\unger{For recent articles,
L.~I.~Unger, Phys.~Rev.~{\bf D48} (1993) 3319. \nextline
W.~Sakuler, W.~Burger, M.~Faber, H.~Marukum, M.~Muller,
P.~De Forcrand, A.~Nakamura and I.~O.~Stamatescu,
Phys.~Lett.~{\bf B276} (1992) 155.
}[\unger].
The screening effect on the long-range correlations
between quarks corresponds to the reduction of the strong correlation
in the infrared momentum region in the gluon propagator.
Hence, the gluon propagator should be modified by the $q$-$\bar q$
pair creation effect.

The dynamical chiral-symmetry breaking is also one of the most
important properties of the light-quark dynamics, and has been studied
by using the Schwinger-Dyson equation
\REF\higashijima{
K.~Higashijima, Phys.~Rev.~{\bf D29} (1984) 1228;
Prog.~Theor.~Phys. Suppl. {\bf 104} (1991) 1.}
\REF\miransky{
V.~A.~Miransky, Sov.~J~.~Nucl.~Phys.~{\bf 38}(2) (1983) 280.
}
\REF\barducci{
A.~Barducci, R.~Casalbuoni, S.~De Curtis,
D.~Dominici and R.~Gatto, Phys.~Lett.~{\bf B193} (1987) 305. \nextline
A.~Barducci, R.~Casalbuoni, S.~De Curtis,
D.~Dominici and R.~Gatto, Phys.~Rev.~{\bf D38} (1988) 238.
}
\REF\kugo{
T.~Kugo and M.~G.~Mitchard, Phys.~Lett. {\bf B286} (1992) 335.
\nextline
K-I.~Aoki, T.~Kugo and M.~G.~Mitchard, Phys.~Lett. {\bf B266} (1991) 467.
\nextline
K-I.~Aoki, M.~Bando, T.~Kugo, M.~G.~Mitchard and H.~Nakatani, \nextline
Prog.~Theor.~Phys. {\bf 84} (1990) 683.
}[\higashijima-\kugo].
However, most authors used a free gluon propagator in the
Schwinger-Dyson equation, and neglected
the nonperturbative effect related to the color confinement.
In the dual Ginzburg-Landau theory,
one can incorporate the nonperturbative effect related to
the color confinement in the infrared region
to the Schwinger-Dyson equation
by using the gluon propagator including
the effects of QCD-monopole condensation.

In the dual Ginzburg-Landau theory, the energy region is naturally
divided into two kinds of regions corresponding to the perturbative
QCD and the nonperturbative QCD.
In the ultraviolet region, the perturbative features are expected
and the system is described by the use of
the running coupling constant at the one or two loop level in the
perturbative scheme.
On the other hand,  the system should be nonperturbative in the
infrared region and such a nonperturbative effect is included in
the dual Ginzburg-Landau theory as the appearance of
QCD-monopole condensation or the dual Meissner effect.
Hence, one can study the nonperturbative effect related to the color
confinement on the dynamical chiral-symmetry breaking in the dual
Ginzburg-Landau theory.

The light-quark confinement can be also investigated
by using the Schwinger-Dyson equation,
although this issue is rather difficult because
the discussion of the {\rm static} quark potential cannot be
applied to the light quarks unlike the heavy quarks.
Instead, the light-quark confinement is characterized by the
disappearance of the physical poles in the quark propagator.
By using the Schwinger-Dyson equation, one gets the
dynamical quark mass $M(p^2)$ depending on the momentum in the
space-like region.
One can examine the disappearance of the physical poles
in the propagator by the analytic continuation of $M(p^2)$
from the space-like region to the time-like region.

In this paper, we study the nonperturbative features of QCD,
the color confinement, the $q$-$\bar q$ pair creation and the
dynamical chiral-symmetry breaking by using the dual
Ginzburg-Landau theory as a phenomenological theory based on QCD,
which includes the mechanism of QCD-monopole condensation
and the dual Meissner effect.
We study their mutual relations and
construct the unified picture of these important features of
the nonperturbative QCD in the infrared region
and the perturbative QCD in the ultraviolet region
in terms of the dual Ginzburg-Landau theory.

In chapter 2, we review the relation between the abelian gauge fixing
and the appearance of the QCD-monopole in line with
't~Hooft's study [\thooftB]. By the abelian gauge fixing, the
nonabelian gauge group is reduced to the abelian one, and abelian
monopoles appear as singularities of the residual abelian gauge field.

In chapter 3, we construct the dual Ginzburg-landau theory
in line with the Kanazawa group [\suzuki,\maedanA]
by introducing the possible self-interaction between abelian
monopoles similar to the Ginzburg-Landau theory in the
superconductivity.

In chapter 4, we derive the static quark potential by investigating
the correlation between the quark currents
within the quenched approximation [\suzuki,\maedanA].
The quark potential has the linear part and the Yukawa part,
and is compared with a phenomenological potential.
We obtain a simple expression for the string tension,
which is analogous to that of the energy per unit length of
the vortex in the superconductivity.

In chapter 5, we study the dynamical effects of the light quarks
on the quark confining potential.
The linear potential is screened in the long-range region
due to the $q$-$\bar q$ pair creation [\unger].
The screening length of the linear potential is estimated
by using the Schwinger formula of the $q$-$\bar q$ pair creation
in the hadron flux tube
\REF\suganumaA{
H.~Suganuma and T.~Tatsumi, Phys.~Lett. {\bf B269} (1991) 371. \nextline
H.~Suganuma and T.~Tatsumi, Prog.~Theor.~Phys. {\bf 90} (1993) 379.
}
\REF\suganumaB{H.~Suganuma and T.~Tatsumi, Ann.~Phys.~(N.Y.)~{\bf 208}
(1991) 470.
}
[\suganumaA, \suganumaB].
By introducing the infrared cutoff in the gluon propagator,
the screening effect for the quark potential is reproduced
in the long-range region.
We obtain a compact formula for the screened quark potential
in the dual Ginzburg-Landau theory.

In chapter 6, we investigate the dynamical chiral-symmetry breaking
by the use of the Schwinger-Dyson equation, where
the gluon propagator includes both the nonperturbative effect
in relation with the confinement and the infrared cutoff
due to the $q$-$\bar q$ pair creation effect.
Since the double pole of the gluon propagator
vanishes due to the infrared cutoff, we can calculate the
Schwinger-Dyson equation without any difficulties of the infrared
divergence.
By solving the Schwinger-Dyson equation within the rainbow
approximation, we find a large enhancement of the dynamical
chiral-symmetry breaking due to QCD-monopole condensation.
The physical quantities related to the chiral-symmetry breaking are
reproduced by the use of the consistent values for
the gauge coupling constant and the QCD scale parameter
with the color confinement and the perturbative QCD.
By extrapolating the quark mass function $M(p^2)$,
which is obtained by the Schwinger-Dyson equation,
to the time-like region, we also examine the existence of the pole
in the light-quark propagator.
We find the disappearance of poles in the light-quark propagator,
which means the confinement of the light quark.

Section 7 is devoted to the summary and discussions.

\chapter{ Abelian Gauge Fixing and QCD-Monopoles }

In this chapter, we review the abelian gauge fixing in the
nonabelian gauge theories and the appearance of the degree
of freedom of the QCD-monopole in line with 't~Hooft's work
[\thooftB].

The color confinement is characterized by the vanishing
of the color dielectric constant and the exclusion of the
color-electric field in the nonperturbative QCD vacuum.
Such features can be explained by the dual Meissner effect
caused by magnetic monopole condensation [\thooftA, \mandelstam],
which is the dual version of Cooper-pair condensation
in the superconductivity.
In this respect, the abelian gauge fixing [\thooftB]
is one of the most interesting gauge for the study of color
confinement,
because the degrees of freedom of the magnetic monopole
naturally appears in this gauge.

The abelian gauge fixing is defined by the diagonalization of
an arbitrary gauge-dependent variable, $X(x)=T^aX^a(x)$,
where $T^a (a=1,2,..,N_c^2-1)$ denotes the generator of the
nonabelian gauge group, SU($N_c$). In this gauge, $X(x)$ is
diagonalized by the gauge transformation, in the fundamental
representation of SU($N_c$),
$$
X(x)\rightarrow X'(x)=\Omega (x)X(x)\Omega ^{-1}(x)
={\rm diag}(\lambda _1(x),\lambda _2(x),..,\lambda _{N_c}(x))
\equiv X_d(x),
\eqn\Xdiag
$$
where
$\Omega (x)=\exp(iT^a\chi ^a(x))\in {\rm SU} (N_c)$ is a gauge function,
Minkowski space.

In the abelian gauge, there remains the abelian gauge
symmetry of [U(1)]$^{N_c-1}$, the maximal torus subgroup of SU($N_c$),
because $X_d(x)$ is invariant under the gauge transformation
defined by a gauge function,
$\omega (x)\in{\rm [U(1)]}^{N_c-1}\subset {\rm SU}(N_c)$, or
$\omega (x)={\rm diag}(e^{i\psi _1(x)}, e^{i\psi _2(x)},.., e^{i\psi
_{N_c}(x)})$
with a constraint $\sum_{a=1}^{N_c}\psi _a=0$,
that is, one finds  $\omega (x)X_d(x)\omega ^{-1}(x)=X_d(x)$.
The off-diagonal elements of the gauge degrees of freedom
are frozen, but the diagonal elements of the gluon field
remain as the gauge degrees of freedom in the abelian gauge.
In other words, nonabelian gauge theories are
reduced to the abelian gauge theories by imposing the
abelian gauge fixing condition, and the off-diagonal gluons
are regarded as charged matter fields in terms of the residual
abelian gauge symmetry.

Next we show the appearance of QCD-monopoles [\thooftB],
magnetic monopoles
in terms of the residual abelian gauge field, in the abelian gauge.
At degeneracy points of the eigenvalues of $X(x)$,
the abelian gauge fixing is not unique even for the off-diagonal
part, and they behave as singular points of the residual
abelian gauge field. Then, these degeneracy points become
monopoles with respect to the abelian gauge field.
Let us examine the above argument by taking the $N_c=2$ case
for simplicity, although the generalization to an arbitrary
$N_c$ case is straightforward.

We investigate the topological nature near the degeneracy points
of the eigenvalues of $X(x)={\tau ^a \over 2}X^a(x)$.
Since two eigenvalues of $X(x)$ are given by
$\pm {1 \over 2} (X_1^2(x)+X_2^2(x)+X_3^2(x))^{1/2}$,
one finds $X_1(x)=X_2(x)=X_3(x)=0$ at the degeneracy points.
Since these three conditions are independent generally,
the degeneracy points form the point-like manifolds in
the three-dimensional space ${\bf R}^3$ at each time $t$.
\foot{
We will consider the system at each $t$, and
omit the time variable $t$ hereafter in this chapter.
}
Let us consider the neighborhood of a degeneracy point
${\bf x}_0\in{\bf R}^3$, which satisfies $X({\bf x}_0)=0$.
By using the Taylor expansion near ${\bf x}_0$, one obtains
$$
X({\bf x})
={\tau ^a \over 2} C^{ai} ({\bf x}-{\bf x}_0)^i, \quad
C^{ai} \equiv \partial^iX^a({\bf x}_0),
\eqn\TAYLOR
$$
where $O(({\bf x}-{\bf x}_0)^2)$ is neglected.
One finds ${\rm det} C \ne 0$ when the manifold of the degeneracy
points is point-like in ${\bf R}^3$, which corresponds to the
general case.

In order to examine the topology of the gauge field
near ${\bf x}_0$, it is useful to introduce the new
space-coordinate variable ${\bf w}$ such that
$w^a\equiv C^{ai}({\bf x}-{\bf x}_0)^i$ instead of ${\bf x}$.
{}From the fact of ${\rm det} C \ne 0$ in the general case,
such a transformation of the coordinate variable is simply
reduced to the linear transformation near the degeneracy point,
so that the topological nature is not changed.
In terms of the new coordinate $w^a$, one finds
the hedgehog configuration,
$$
X(w^a)={\tau ^a \over 2} w^a
\eqn\HEDGE
$$
near the degeneracy point, $w^a=0$, and
the eigenvalues of $X$ is found to be
$\pm {1 \over 2}(w_1^2+w_2^2+w_3^2)^{1/2}$.
This hedgehog configuration {\HEDGE}
is one of the simplest nontrivial solution
corresponding to the homotopy group
$\pi _2({\rm SU}(2)/{\rm U}(1)) = \pi _1({\rm U}(1))= Z_\infty $,
and such a configuration widely appears in the particle physics,
for instance, the 't~Hooft-Polyakov monopole
\REF\thooftC{
G.~'t~Hooft, Nucl.~Phys.~{\bf B79} (1974) 276. \nextline
A.~M.~Polyakov, JETP Lett.~{20} (1974) 194.
}[\thooftC],
the instanton in the nonabelian gauge theory
\REF\rajaraman{
For instance,
R.~Rajaraman, ``Solitons and Instantons",
(North-Holland, Amsterdam, 1982) 1
and references therein.
}[\rajaraman],
and the chiral soliton in the Skyrme-Witten model
\REF\skyrme{
V.G.Makhoankov, Y.~P.~Rybakov and V.~I.~Sanyuk,
``The Skyrme model",
(Springer-Verlag, Berlin Heidelberg, 1993.) 1
and references therein.
}[\skyrme].
It should be noted that such a topological configuration is
originated from the {\it nonabelian} nature.

By the use of the parametrization such as \nextline
${\bf w}=(w_1,w_2,w_3)=(r\cos \phi \sin \theta , r\sin \phi \sin \theta , r\cos
\theta )$,
$X$ can be expressed as
$$
X=
{1 \over 2}
\pmatrix {
w_3      &  w_1-iw_2 \cr
w_1+iw_2 &  -w_3
}
=
{r \over 2}
\pmatrix{
\cos \theta       & e^{-i\phi }\sin \theta  \cr
e^{i\phi }\sin \theta  &  -\cos \theta
}.
\eqn\MATRIX
$$
Hence, the gauge function $\Omega $$\in$ SU(2) which diagonalizes $X$ is
found to be
$$
\Omega ({\bf w})=
\pmatrix{
e^{i\phi }\cos{\theta \over 2}  &   \sin{\theta \over 2} \cr
-\sin{\theta \over 2}        &   e^{-i\phi }\cos{\theta \over 2}
},
\eqn\OMEGA
$$
which leads $\Omega X\Omega ^{-1}=X_d$.
By the gauge transformation $\Omega $, the gauge field is transformed as
$$
A_\mu \rightarrow A_\mu '=\Omega (A_\mu -{i \over e}\partial_\mu )\Omega ^{-1},
\eqn\GAUGETR
$$
where $\partial_\mu  \equiv {\partial \over \partial x^\mu }$.
Since the original gauge field $A_\mu $ is regular,
$\Omega A_\mu \Omega ^{-1}$ is also regular.
On the other hand, $A_\mu ^s \equiv -{i \over e}\Omega \partial_\mu \Omega
^{-1}$
becomes singular, as is shown in the following.
Consider the line-integral of $A_\mu ^s$ along a contour $C$,
$$
\eqalign{
 \Phi     &\equiv-\int_C A_\mu ^sdx^\mu
       =\int_C A_i^sdx_i
       =-{i \over e}\int_C \Omega \partial_i \Omega ^{-1}dx_i \cr
       &=-{i \over e}\int_C
          \Omega {\partial \over \partial w_a} \Omega ^{-1}dw_a
       =\int_C {\cal A}_a^s({\bf w})dw_a,
}
\eqn\FLUXa
$$
where ${\cal A}_a\equiv -{i \over e}
\Omega ({\bf w}){\partial \over \partial w_a}\Omega ^{-1}({\bf w})$.
When $C$ is a closed loop,
$\Phi $ is identified as the magnetic flux which penetrates the area
inside the closed contour $C$ by way of the Stokes theorem.

Let $C$ be taken as a closed circle such that $r$ and $\theta $
are constant, while
\nextline
$\phi  \in [0,~2\pi )$, for simplicity. Then, one gets
$$
\eqalign{
\Phi (\theta )&=
-{i \over e}\int_0^{2\pi }d\phi \Omega {\partial \over \partial \phi }\Omega
^{-1}\cr
    &=-{1 \over e}\int_0^{2\pi }d\phi
\pmatrix{\cos^2{\theta  \over 2}  &
    -e^{i\phi }\sin{\theta  \over 2} \cos{\theta  \over 2} \cr
    -e^{-i\phi }\sin{\theta  \over 2} \cos{\theta  \over 2} &
    -\cos^2{\theta  \over 2}
    }\cr
  &=-{2\pi  \over e}
\pmatrix{\cos^2{\theta  \over 2}  &   0                \cr
    0                        &   -\cos^2{\theta  \over 2}
    }
  =-{2\pi  \over e}\cos^2{\theta  \over 2}\cdot \tau _3
  =-{2\pi  \over e}(1+\cos \theta ){\tau _3 \over 2}.
}
\eqn\FLUXb
$$
Since the final formula of $\Phi $ is diagonal, $\Phi $ is identified
as the magnetic flux of the abelian gauge field.

It is notable that $\Phi $ is finite even at $\theta $=0, where contour
$C$ shrinks into a point. This means that the corresponding
magnetic field diverges at $\theta $=0 like a $\delta $-function,
and the gauge field is singular at $\theta $=0.
Then, the magnetic flux is decomposed into two parts,
$$
\Phi (\theta )=-{2\pi  \over e}(1+\cos \theta ){\tau _3 \over 2}
  ={4\pi  \over e} \{ {1-\cos \theta  \over 2}-1 \} {\tau _3 \over 2},
\eqn\FLUXc
$$
where the first and the second terms in the curly bracket
are nothing but the expressions of the magnetic flux of the
magnetic monopole with the magnetic charge $g={4\pi  \over e}$
and the corresponding Dirac string, respectively.
The relation $eg=4\pi $ corresponds to the Dirac condition
for magnetic monopoles [\rajaraman].
Hence, the abelian gauge field near the degeneracy point of
eigenvalues of $X$ is equivalent to the magnetic monopole
system in the $w$-coordinate space.

Since the topological natures are not changed by the
transformation between $x$ and $w$, the degeneracy points
also become the magnetic monopoles in the abelian gauge field
in terms of the $x$-coordinate space.
Thus, the QCD-monopoles appear at the degeneracy point of
the eigenvalues of $X$ in the abelian gauge.
In this scheme, the degeneracy condition ${\rm det} X=0$
is independent of the representation of the nonabelian group.
Hence, the number and the location of the abelian magnetic
monopole do not depend on the representation for each $X$,
although this gauge fixing depends on the
selection of the gauge dependent variable $X$.

Thus, nonabelian gauge theories are reduced to abelian gauge
theories with QCD-monopoles in the abelian gauge,
and the degrees of freedom of QCD-monopoles naturally
appear in QCD in this gauge.
If the QCD-monopole is condensed like the Cooper pair in the
superconductivity theory, one expects the dual Meissner effect
[\thooftA-\thooftB],
the vanishing of the color dielectric constant and the exclusion
of the color-electric flux, which characterizes the color confinement.

In recent years, several authors found evidences
of the abelian dominance, that is
the relevant roles of the abelian degrees of freedom to the
nonperturbative quantities in the nonabelian gauge theory
by using the lattice gauge simulation [\yotsuyanagi,\hioki].
They pointed out that the abelian gauge configurations
seems to play a dominant role to the nonperturbative quantities
like the Wilson loop or the Polyakov loop
in the lattice simulations of the nonabelian gauge theory
[\yotsuyanagi,\hioki].
The abelian dominance gives a consistent picture with
't~Hooft's conjecture, the relevance of the abelian gauge fixing
to the confinement.

The role of the QCD-monopole to the confinement
has been also examined by the use of the lattice gauge
simulations. The compact QED [\daggoto] has a confining phase
in the strong-coupling region as well as QCD,
and the confinement is known to be realized due to
monopole condensation [\banks, \degrand].
The importance of the QCD-monopole to the confinement
has been also reported in the nonabelian gauge theories.
Some evidences of QCD-monopole condensation
was reported by Kronfeld et al. [\kronfeld] by comparing
the SU(2) gauge theory and the compact QED
in the lattice gauge theories.
Another group has reported a drastic change of the space-time
asymmetry of the monopole currents at the deconfinement phase
transition, which may provide an interesting correspondence between
the spatial correlation of the monopoles
and the color confinement [\hioki].
Although there is no difference between the time component and the
space one of the current in the confining phase,
there is a clear enhancement of the time component of the
monopole current in the deconfinement phase [\hioki].
This change suggests that the QCD-monopoles
become more static in the deconfinement phase,
which can be regarded as a reduction of the spatial correlation
of the monopoles.
Very recently, the Kanazawa group investigated
the energy and the entropy of a monopole loop
by using the SU(2) lattice gauge simulation [\shiba].
They found that the entropy of an extended monopole
loop dominates over its energy and monopole
condensation takes place on the renormalized lattice
in the infrared region, $\beta <\beta _c$ [\shiba].
They also showed that the string tension of the SU(2) gauge theory
are well reproduced by such a extended monopole contributions alone
[\shiba].
Thus, recent lattice gauge simulations have given
the phenomenological evidences of the abelian dominance
and the important role of the QCD-monopole to the color
confinement. Hence, 't~Hooft's conjecture of QCD-monopole
condensation has been accepted as a realistic interpretation
of the color confinement in the infrared region.

Finally, we reconsider the role of the nonabelian nature
to the appearance of QCD-monopoles in this scheme.
It is notable that the off-diagonal part of the gauge field
contributes to the QCD-monopole or the hedgehog configuration
{\HEDGE} as well as the diagonal part, and the
corresponding gauge configuration $A_\mu '(x)$ in Eq.{\GAUGETR}
has both the diagonal and the off-diagonal parts,
although the color-magnetic flux $\Phi $ in Eq.{\FLUXb} is diagonal.
In this respect, these monopoles are regarded as the collective
modes including the effect of the off-diagonal gauge field.
In the topological point of view,
the nonabelian nature of the nonabelian gauge theory
plays an essential role on the appearance of the hedgehog
configuration {\HEDGE} in the full gauge space corresponding
to the nontrivial homotopy group
$\pi _2({\rm SU}(N_c)/{\rm [U(1)]}^{N_c-1})
=\pi _1({\rm [U(1)]}^{N_c-1})=Z_\infty ^{N_c-1} $
[\rajaraman],
and this configuration behaves as the singularity or the Dirac monopole
in terms of the residual abelian gauge field in the abelian gauge.

It is interesting to compare the superconductivity with
the nonabelian gauge system in the abelian gauge.
In an ordinary sense,
the pure-gauge system seems quite different
from the complicated matter system in the superconductor
including the electron and the metallic lattice.
However, even the pure-gauge system of the nonabelian gauge theory
may give a similar situation
to the superconductor by using the abelian gauge fixing.
In the abelian gauge,
the diagonal part and the off-diagonal part of the gauge field
come to play a different role in a physical meaning.
The diagonal gauge fields remain to be gauge fields,
while the off-diagonal gauge fields behave as
{\it charged matter fields }
and permit the appearance of the QCD-monopoles
as mentioned above.
Then, this off-diagonal gauge fields in the abelian gauge
may play a similar role to the {\it matter system}
in the superconductor,
and therefore one may expect the similar situation to the
complicated matter system of the superconductor.

In the next chapter, we construct the phenomenological theory
based on QCD including the dual Meissner effect, and
study the nonperturbative phenomena of QCD, not only the
confinement but also the $q$-$\bar q$ pair creation and
the dynamical chiral-symmetry breaking, based on 't~Hooft's
conjecture, the relevance of the abelian gauge and the abelian
monopoles in the infrared region.

\chapter{ The Dual Ginzburg-Landau Theory and QCD-Monopole
Condensation }

In this chapter, we construct the effective theory
of the nonperturbative QCD in the abelian gauge,
by introducing the QCD-monopole field
as a relevant mode responsible to the color confinement.
The QCD Lagrangian is described by quarks $q^\alpha $ and gluons $A_\mu $
[\nachtmann],
$$
{\cal L}_{\rm QCD}
=-{1 \over 4}G_{\mu \nu }^aG^{\mu \nu }_a+\bar q (i\not D-m)q,
\eqn\QCDa
$$
where $G_{\mu \nu } \equiv \partial_\mu  A_\nu -\partial_\nu  A_\mu +ie[A_\mu
,A_\nu ]$
and $D_\mu \equiv \partial_\mu +ieA_\mu $ are the field strength tensor and
the covariant derivative, respectively.

We shall take the abelian gauge fixing presented by 't~Hooft.
There remains the gauge symmetry in the diagonal part of
the gauge degrees of freedom in the abelian gauge, so that
the diagonal gluons would bring a large contribution to the
dynamical process as the gauge fields even in this gauge.
On the contrary, off-diagonal gluons behave as
charged matter fields instead of the gauge fields
because the off-diagonal parts of the gauge degrees of freedom
are fixed in this gauge [\thooftB].
The appearance of QCD-monopoles is important and
should be taken into account in the abelian gauge.
We consider the main role of the off-diagonal gluon is to form
the QCD-monopoles in this gauge [\suzuki],
which was supported by the recent Monte Carlo simulation
[\yotsuyanagi,\hioki,\shiba].
The infrared effective theory of QCD is thus constructed by
the abelian gauge fields and the QCD-monopole fields in the
abelian gauge.

To begin with, let us consider the abelian gauge part in
the effective theory,
$$
{\cal L}_{\rm Abel}
  \equiv -{1 \over 4}\vec f_{\mu \nu }^2
   +\bar q (i\not \partial-e \not \vec A \cdot \vec H-m)q
\eqn\QCDb
$$
where $\vec f_{\mu \nu }\equiv \partial_\mu  \vec A_\nu -\partial_\nu \vec
A_\mu $,
$\vec H \equiv(T_3, T_8)$, and $\vec A^\mu  \equiv (A^\mu _3, A^\mu _8)$.
Owing to the duality of the abelian gauge theory,
one can introduce the QCD-monopole field and study the dual
Meissner effect by the parallel argument to the superconductivity.
To make the duality of the gauge theory manifest,
it is convenient to use the Zwanziger form
\REF\zwanziger{
D.~Zwanziger, Phys.~Rev.~{\bf D3} (1971) 880.
}[\zwanziger],
$$
\eqalign{
-{1 \over 4}\vec f_{\mu \nu }^2
&=-{1 \over 4}(\partial_\mu \vec A_\nu -\partial_\nu  \vec A_\mu )^2
=-{1 \over 4}(\partial_\mu \vec B_\nu -\partial_\nu  \vec B_\mu )^2 \cr
&=-{1 \over 2n^2}
[n\cdot (\partial \wedge \vec A)]^\nu
[n\cdot ^*(\partial \wedge \vec B)]_\nu
+{1 \over 2n^2}
[n\cdot (\partial \wedge \vec B)]^\nu
[n\cdot ^*(\partial \wedge \vec A)]_\nu  \cr
&-{1 \over 2n^2}
[n\cdot (\partial \wedge \vec A)]^2
-{1 \over 2n^2}
[n\cdot (\partial \wedge \vec B)]^2,
}
\eqn\ABELa
$$
where $n_\mu $ is an arbitrary constant four vector,
corresponding to the direction of the Dirac string,
$[a\cdot (b\wedge c)]^\nu \equiv a_\mu (b^\mu c^\nu -b^\nu c^\mu )$, and
$[a\cdot ^*(b\wedge c)]^\nu \equiv a_\mu \epsilon ^{\mu \nu \alpha \beta
}(b_\alpha c_\beta )$.
The field denoted by $\vec B_\mu $ is called the dual gauge field,
because it satisfies the relation
$$
\ ^* \vec f_{\mu \nu }=\partial_\mu  \vec B_\nu -\partial_\nu  \vec B_\mu
\eqn\DUALa
$$
in the absence of the electric current.
Then, the Lorentz indices of the electric and magnetic fields
are interchanged each other by using the dual gauge field
$\vec B_\mu $, instead of $\vec A_\mu $ in this case.

In the Zwanziger form, the magnetic current $\vec k_\mu $ directly
couples with the dual gauge field $\vec B_\mu $
[\zwanziger], similar to the
coupling between the electric current $\vec j_\mu $
and the ordinary gauge field $\vec A_\mu $.
Hence, the interaction term in the presence of the electric current
and the magnetic current is given by
${\cal L}_{\rm int}=\vec j_\mu  \vec A^\mu +\vec k_\mu  \vec B^\mu $.

It is notable that the last expression in Eq.{\ABELa}
is invariant under the two types of the local transformations,
$$
\eqalign{
\vec A_\mu (x)&\rightarrow \vec A_\mu (x)+{1 \over e}\partial_\mu \vec \theta
_A(x), \cr
\vec B_\mu (x)&\rightarrow \vec B_\mu (x)+{1 \over g}\partial_\mu \vec \theta
_B(x),
}
\eqn\SYMa
$$
where $\theta _A(x)$ and $\theta _B(x)$ are independent arbitrary scalar
functions. Then, we see the extended local symmetry,
$[{\rm U}(1)_e]^2\times [{\rm U(1)}_m]^2$ in the Zwanziger form.
However, it does not mean the increase of the gauge degrees of
freedom, because $\vec A_\mu $ and $\vec B_\mu $ are not independent
fields due to the interaction between them in the Zwanziger
form {\ABELa}
\REF\blagojevic{
M.~Blagojevic and P.~Senjanovic, Nucl.~Phys.~{\bf B161} (1979) 112.
}
[\zwanziger, \blagojevic].

We phenomenologically introduce the QCD-monopoles as the relevant
degrees of freedom to the color confinement
in order to investigate the nonperturbative features in QCD.
By introducing the QCD-monopole field $\chi _\alpha $($\alpha $=1,2,3) and
its coupling with the dual gauge fields $\vec B_\mu $
\REF\bardacki{
K.~Bardacki and S.~Samuel,  Phys.~Rev.~{\bf D18} (1978) 2849.\nextline
S.~Samuel,  Nucl.~Phys.~{\bf B154} (1979) 62.
}
[\bardacki],
the effective Lagrangian is obtained by
$$
\eqalign{
{\cal L}_{\rm DGL}
&=-{1 \over 2n^2}
[n\cdot (\partial \wedge \vec A)]^\nu
[n\cdot ^*(\partial \wedge \vec B)]_\nu
+{1 \over 2n^2}
[n\cdot (\partial \wedge \vec B)]^\nu
[n\cdot ^*(\partial \wedge \vec A)]_\nu  \cr
&-{1 \over 2n^2}
[n\cdot (\partial \wedge \vec A)]^2
-{1 \over 2n^2}
[n\cdot (\partial \wedge \vec B)]^2 \cr
&+\bar q (i\not \partial-e \not \vec A \cdot \vec H-m)q
+\sum_{\alpha =1}^3[|(i\partial_\mu -g \vec \epsilon _\alpha  \cdot \vec B_\mu
)\chi _\alpha |^2
-\lambda (|\chi _\alpha |^2-v^2)^2],
}
\eqn\ZWANa
$$
where the self-interaction of $\chi _\alpha $ is introduced
phenomenologically similar to the Ginzburg-Landau theory in the
superconductivity physics
[\suzuki,\maedanA].
We call the theory based on the Lagrangian {\ZWANa} as the
dual Ginzburg-Landau theory because it is a dual version of the
Ginzburg-Landau theory.

In Eq.{\ZWANa},
$g$ is the unit magnetic charge of the magnetic monopoles,
obeying the Dirac condition, $eg=4\pi $.
Let us consider the QCD-monopoles belonging to the fundamental
representation of SU(3)$_c$.
In this case, the magnetic charge $\vec \epsilon _\alpha $ ($\alpha $=1,2,3)
of the QCD-monopole field $\chi _\alpha $ is found to be
$\vec \epsilon _1=(1, 0)$, $\vec \epsilon _2=(-{1 \over 2}, -{\sqrt{3} \over
2})$,
and $\vec \epsilon _3=(-{1 \over 2}, {\sqrt{3} \over 2})$
in a suitable representation of SU(3)$_c$ [\suzuki,\maedanA].
In the Lagrangian {\ZWANa}, three kinds of the QCD-monopole fields
$\chi _\alpha $ ($\alpha $=1,2,3) are complex fields, and there is a constraint
among their phases, e.g. $\sum_{\alpha =1}^3 {\rm arg} \chi _\alpha =0$
[\suzuki,\maedanA].
One finds that the residual two degrees of freedom on the
phase of $\chi _\alpha $ are closely related with the two kinds of the
dual gauge field $\vec B^\mu =(B^\mu _3, B^\mu _8)$.
The dual gauge symmetry
$[{\rm U(1)}_m]^2 \equiv {\rm U(1)}_m^{(3)} \times {\rm U(1)}_m^{(8)}$
in {\SYMa} is connected to the local phase transformation of
the QCD-monopole field,
$$
\eqalign{
\vec B_\mu (x)&\rightarrow \vec B_\mu (x)+{1 \over g}\partial_\mu \vec \theta
_B(x), \cr
\chi _\alpha (x)&\rightarrow e^{-i \vec \epsilon _\alpha  \cdot \vec \theta
_B(x)} \chi _\alpha (x)
}
\eqn\SYMb
$$
similar to the ordinary gauge symmetry
$[{\rm U(1)}_e]^2 \equiv {\rm U(1)}_e^{(3)} \times {\rm U(1)}_e^{(8)}$.

The dual Meissner effect is brought by the realization of
QCD-monopole condensation due to the self-interaction
of $\chi _\alpha $ with $v^2>0$ in the Lagrangian {\ZWANa}.
QCD-monopole condensation leads the mass term of the
dual gauge field  $\vec B_\mu $, similar to the generation of the photon
mass due to Cooper-pair condensation in the superconductivity.
For the QCD vacuum, one finds  $|\chi _\alpha |=v (\alpha =1,2,3)$
in the mean field level of the $\chi $-field, and therefore
the Lagrangian {\ZWANa} reads
$$
\eqalign{
{\cal L}_{\rm DGL-MF}
&=-{1 \over 2n^2}
[n\cdot (\partial \wedge \vec A)]^\nu
[n\cdot ^*(\partial \wedge \vec B)]_\nu
+{1 \over 2n^2}
[n\cdot (\partial \wedge \vec B)]^\nu
[n\cdot ^*(\partial \wedge \vec A)]_\nu  \cr
-{1 \over 2n^2}&
[n\cdot (\partial \wedge \vec A)]^2
-{1 \over 2n^2}
[n\cdot (\partial \wedge \vec B)]^2
+\bar q (i\not \partial-e \not \vec A \cdot \vec H-m)q
+{1 \over 2}m_B^2 \vec B^2,
}
\eqn\ZWANb
$$
where $m_B=\sqrt 3 gv$ is the mass of the dual gauge field
$\vec B_\mu $.
The QCD-monopole field is also massive due to the
self-interaction, $m_\chi =2\sqrt{\lambda }v$.
It is notable that the dual gauge symmetry $[{\rm U}(1)_m]^2$
in {\SYMa} is broken, while there remains the gauge symmetry
$[{\rm U}(1)_e]^2$, which is the subgroup of the original
gauge group ${\rm SU}(N_c)$.
Hence, QCD-monopole condensation never breaks
the abelian gauge symmetry $[{\rm U}(1)_e]^2$, subgroup of SU(3)$_c$.
\foot{
QCD-monopole condensation may break the
off-diagonal part of the original gauge symmetry SU(3)$_c$.
It is, however, less important because the off-diagonal part
already has no gauge symmetry due to the abelian gauge fixing.
}

To investigate the QCD vacuum, it is useful to start with the
mean-field level Lagrangian {\ZWANb}, neglecting the fluctuation
of the QCD-monopole field.
By integrating out the dual gauge field $\vec B_\mu $
in the partition functional, one gets the Lagrangian
described by the gauge field $\vec A_\mu $,
$$
\eqalign{
{\cal L}_{\rm DGL-MF}
  &=-{1 \over 4}\vec f_{\mu \nu } \vec f^{\mu \nu }
   +{1 \over 2}\vec A^\mu  K_{\mu \nu } \vec A^\nu
   +\bar q (i\not \partial-e \not \vec A \cdot \vec H-m)q.
}
\eqn\ABELb
$$
In this equation, the non-local operator $K^{\mu \nu }$ is defined by
$$
K^{\mu \nu }
\equiv {n^2 m_B^2 \over (n \cdot \partial)^2 + n^2 m_B^2 }X^{\mu \nu },
\eqn\Kdef
$$
where $X^{\mu \nu }$ is given by
$$
\eqalign{
X^{\mu \nu }&\equiv{1 \over n^2}\epsilon _\lambda  \ ^{\mu \alpha \beta
}\epsilon ^{\lambda \nu \gamma \delta }
n_\alpha n_\gamma \partial_\beta \partial_\delta \cr
&={1 \over n^2}[-n^2\partial^2g^{\mu \nu }+(n\cdot\partial)^2g^{\mu \nu }
+n^\mu n^\nu \partial^2-(n\cdot\partial)(n^\mu \partial^\nu +n^\nu \partial^\mu
)
+n^2\partial^\mu \partial^\nu ].
}
\eqn\Xdef
$$

One easily finds the $[{\rm U}(1)_e]^2$-gauge invariance of the Lagrangian
${\cal L}_{\rm DGL-MF}$ in Eq.{\ABELb} by using the relations,
$$
X_{\mu \nu }=X_{\nu \mu }, \quad  X_{\mu \nu }\partial^\nu =X_{\mu \nu }n^\nu
=0.
\eqn\Xa
$$
Hence, we find again that QCD-monopole condensation never breaks the
gauge symmetry $[{\rm U}(1)_e]^2\in {\rm SU}(3)$.
The $m_B$-dependent term in the effective Lagrangian {\ABELb}
originally corresponds to the mass term of the dual
gauge field $\vec B_\mu $, and leads to the dual Meissner effect
or the color confinement.

We discuss the spatial distribution of the QCD-monopole condensate
$|\chi _\alpha (x)|$ and the color-electric field $E(x)$ in the hadron flux
tube within the quenched approximation [\maedanB].
We consider the flux tube in the mesonic system
including the static quark and the antiquark
with a relative distance ${\bf r}$.
A straight flux tube is formed between the quark and
the antiquark, and therefore the system is axial symmetric
around the axis between them [\maedanB].
In a large separation limit, $|{\bf r}|\rightarrow \infty $, the system has a
translational invariance along the direction
$\hat {\bf r}$, and
the system becomes quite similar to a vortex solution in the
superconductivity
\REF\lifshitz{
E.~M.~Lifshitz and L.~P.~Pitaevsii,
Vol.9 of Course of Theoretical Physics,
``Statistical Physics, Part 2",
(Pergamon press, Oxford, 1981) 1.
}[\lifshitz] or
the Nielsen-Olesen vortex in the abelian Higgs model
\REF\nielsen{
H.~B.~Nielsen and P.~Olesen, Nucl.~Phys.{\bf B61} (1973) 45.
}
[\nielsen].

In this subject, it is convenient to describe the
dual Ginzburg-Landau Lagrangian {\ZWANa} only by the $\vec B_\mu $ field,
$$
\eqalign{
{\cal L}_{\rm DGL}=
-{1 \over 4}(\partial_\mu  \vec B_\nu - \partial_\nu  \vec B_\mu  )^2
+\sum_{\alpha =1}^3[|(i\partial_\mu -g \vec \epsilon _\alpha \cdot \vec B_\mu
)\chi _\alpha |^2
-\lambda (|\chi _\alpha |^2-v^2)^2].
}
\eqn\VORTa
$$
We investigate the solution of the coupled equation of the abelian
monopole field $|\chi _\alpha (x)|$ and the $\vec B_\mu (x)$ field
at the tree level, which is analogous to the vortex solution
in the superconductivity [\lifshitz].
The solutions for the color-electric field and
the QCD-monopole field are given by
functions of $\rho $, the distance from the
cylindrical axis, and take forms as
$E_{\rm diag}(\rho )\equiv E_3(\rho )=E_8(\rho )$ and
$|\chi (\rho )|\equiv |\chi _1(\rho )|=|\chi _2(\rho )|=|\chi _3(\rho )|$
in the flux tube [\maedanB].

We consider the case of $m_\chi >m_B$ corresponding to the type II
superconductor because $m_\chi >m_B$ is suggested from the study of
the quark potential as will be shown in the following chapter.
The color-electric field $E_{\rm diag}(\rho )$ takes a large value
only in a region of $\rho  \lsim m_B^{-1}$,
and therefore the cylindrical radius of the hadron flux tube
is roughly given by $m_B^{-1}$.
One finds the reduction of the QCD-monopole condensate $|\chi (\rho )|$
in the central region of $\rho  \lsim m_\chi ^{-1}$ in the flux tube
[\lifshitz,\nielsen].
The QCD-monopole condensate is regarded as an almost constant
value $v$, $|\chi (\rho )|\simeq v$,
for the infrared region $\rho  \gsim m_\chi ^{-1}$ corresponding to
$k_{_T} \lsim  m_\chi $ in the momentum space,
where $k_{_T}$ denotes the transverse momentum component.
On the contrary, the QCD-monopole condensate almost disappear,
$|\chi (\rho )|\simeq 0$ for the ultraviolet region $\rho  \lsim m_\chi ^{-1}$
corresponding to $k_{_T} \gsim m_\chi $ in the momentum space.
Thus, one finds the approximate relations,
$$
|\chi (\rho )|\simeq v\theta (\rho  - m_\chi ^{-1}) \quad
{\rm and } \quad
m_B(\rho )\simeq \sqrt{3} g v \cdot \theta (\rho  - m_\chi ^{-1}),
\eqn\VORTc
$$
and therefore the ultraviolet cutoff appears in $m_B$.
The appearance of the ultraviolet cutoff is also found
in the argument of the vortex in the superconductivity [\lifshitz],
and is important for the study of the string tension
as will be shown in the next chapter.

We have obtained a phenomenological theory of
the nonperturbative QCD, the dual Ginzburg-Landau theory,
which contains the confining mechanism
\nextline
through QCD-monopole condensation and the dual Meissner effect.
By using the Lagrangian {\ZWANa} or {\ABELb} in this theory,
we investigate the nonperturbative features of QCD,
the color confinement, the effect of the $q$-$\bar q$ pair creation,
and the dynamical chiral-symmetry breaking.

\chapter{ Quark Confinement Potential in the Quenched Approximation }

In this chapter, we derive the quark static potential in the dual
Ginzburg-Landau theory, based on the Lagrangian {\ZWANa} or {\ABELb}
within the quenched approximation. The effective Lagrangian
including the quark current $\vec j_\mu $ is given by
$$
\eqalign{
{\cal L}_{\rm DGL-MF}&={1 \over 2}\vec A^\mu D_{\mu \nu }^{-1}\vec A^\nu
+\vec j_\mu  \vec A^\mu \cr
&={1 \over 2} (\vec A^\mu +\vec j_\alpha D^{\alpha \mu })D^{-1}_{\mu \nu }
(\vec A^\nu +D^{\nu \beta }\vec j_\beta )-{1 \over 2}\vec j_\mu D^{\mu \nu
}\vec j_\nu ,
}
\eqn\LPa
$$
where $D_{\mu \nu }$ is the propagator of the diagonal gluon field,
$\vec A_\mu $. By integrating out $\vec A_\mu $ in the partition
functional, one gets the quark-current correlation,
$$
{\cal L}_j=-{1 \over 2}\vec j_\mu D^{\mu \nu }\vec j_\nu .
\eqn\LPb
$$
The diagonal-gluon propagator $D^{\mu \nu }$ is obtained from
the effective Lagrangian {\ABELb} in the dual Ginzburg-Landau theory,
and the inverse propagator of the diagonal gluon field is
given as
$$
D_{\mu \nu }^{-1}=g_{\mu \nu }\partial^2-(1-{1 \over \alpha _e})
\partial_\mu  \partial_\nu
+{m_B^2n^2 \over (n\cdot \partial)^2+m_B^2n^2}X_{\mu \nu }
\eqn\LPc
$$
in the Lorentz gauge with the gauge fixing term,
$
{\cal L}_{G.F.}=-{1 \over 2\alpha _e}(\partial_\mu  \vec A^\mu )^2.
$
Then, the diagonal-gluon propagator is given by
$$
D_{\mu \nu }={1 \over \partial^2}\{g_{\mu \nu }+(\alpha _e-1)
{\partial_\mu  \partial_\nu  \over \partial^2} \}
-{1 \over \partial^2}
{m_B^2 \over \partial^2+m_B^2}
{n^2 \over (n\cdot \partial)^2}X_{\mu \nu },
\eqn\LPd
$$
and therefore the quark-current correlation {\LPb} leads
$$
\eqalign{
{\cal L}_j&=
-{1 \over 2}\vec j^\mu
[{1 \over \partial^2}g_{\mu \nu }
-{1 \over \partial^2}
{m_B^2 \over \partial^2+m_B^2}
{n^2 \over (n\cdot \partial)^2}
\{-\partial^2(g_{\mu \nu }-{n_\mu n_\nu  \over n^2})
+{(n \cdot \partial)^2 \over n^2}g_{\mu \nu }\}]
\vec j^\nu  \cr
&=-{1 \over 2}\vec j^\mu
[{1 \over \partial^2+m_B^2}g_{\mu \nu }
+{m_B^2 \over \partial^2+m_B^2}
{n^2 \over (n\cdot \partial)^2}
(g_{\mu \nu }-{n_\mu n_\nu  \over n^2})]
\vec j^\nu
}
\eqn\LPe
$$
independent of the gauge-fixing parameter $\alpha _e$.
The action including the quark-current correlation is given by
$$
\eqalign{
S_j
& \equiv\int d^4x {\cal L}_j \cr
& =\int{d^4k \over(2\pi )^4}{1 \over 2} \vec j^\mu (-k)
[{1 \over k^2-m_B^2}g_{\mu \nu }
+{-m_B^2 \over k^2-m_B^2}{n^2 \over (n \cdot k)^2}
(g_{\mu \nu }-{n_\mu n_\nu \over n^2})] \vec j^\nu (k),
}
\eqn\JJa
$$
where $\vec j_\mu (k)$ is the Fourier component of $\vec j_\mu (x)$
with $\vec j_\mu (k) \equiv \int d^4x e^{ik \cdot x} \vec j_\mu (x)$.

Let us consider the static system of a heavy quark and
antiquark pair with opposite color-charge located
at {\bf a} and {\bf b}, respectively [\suzuki,\maedanA].
In this case, the quark current is given by
$$
\vec j_\mu (x)=\vec Qg_{\mu 0}\{\delta ^3({\bf x}-{\bf b})
-\delta ^3({\bf x}-{\bf a})\},
\eqn\QCa
$$
where $\vec Q=(Q_3, Q_8)$ is the color charge of the static quark
\REF\kerson{
K.~Huang, ``Quarks, Leptons and Gauge Fields",
World Scientific, \nextline Singapore (1982) 1.}
[\kerson],
and its Fourier component reads
$$
\vec j_\mu (k)=\vec Qg_{\mu 0}2\pi \delta (k_0)
(e^{-i {\bf k} \cdot {\bf b}}-e^{-i {\bf k} \cdot {\bf a}}).
\eqn\QCb
$$
The action {\JJa} reads
$$
\eqalign{
S_j
&=-\vec Q^2 \int dt
\int {d^3k \over (2 \pi)^3}
{1 \over 2}(1-e^{i{\bf k}\cdot {\bf r}})
(1-e^{-i{\bf k}\cdot {\bf r}})[{1 \over {\bf k}^2+m_B^2}
+{m_B^2 \over {\bf k}^2+m_B^2}\cdot {1 \over ({\bf n} \cdot {\bf k})^2}],
}
\eqn\JJb
$$
where ${\bf n}$ is a unit vector and
${\bf r} \equiv {\bf b}-{\bf a}$ is the relative distance
between the quark and the antiquark.
The static quark potential thus obtained is divided
into two parts,
$$
V({\bf r} ; {\bf n})
=V_{\rm Yukawa}(r)+V_{\rm linear}({\bf r} ; {\bf n}).
\eqn\JJc
$$
Here, $V_{\rm Yukawa}(r)$ gives the Yukawa-type potential,
$$
\eqalign{
V_{\rm Yukawa}(r)
&\equiv
\vec Q^2 \int {d^3k \over (2\pi )^3}{1 \over 2}
(1-e^{i{\bf k}\cdot {\bf r}})(1-e^{-i{\bf k}\cdot {\bf r}})
{1 \over {\bf k}^2+m_B^2}\cr
&=-\vec Q^2 \int {d^3k \over (2\pi )^3}
e^{i{\bf k}\cdot {\bf r}}{1 \over {\bf k}^2+m_B^2}
=-{\vec Q^2 \over 4\pi }\cdot {e^{-m_Br} \over r},
}
\eqn\JJd
$$
apart from a irrelevant constant.
$V_{\rm linear}({\bf r} ; {\bf n})$ is given by
$$
\eqalign{
V_{\rm linear}({\bf r} ; {\bf n})
&\equiv
\vec Q^2 \int {d^3k \over (2\pi )^3}{1 \over 2}
(1-e^{i{\bf k}\cdot {\bf r}})(1-e^{-i{\bf k}\cdot {\bf r}})
{m_B^2 \over {\bf k}^2+m_B^2}\cdot {1 \over ({\bf n} \cdot {\bf k})^2}\cr
&=\vec Q^2 \int {d^3k \over (2\pi )^3}
\{1-\cos({\bf k}\cdot {\bf r})\}{m_B^2 \over {\bf k}^2+m_B^2}\cdot
{1 \over ({\bf n} \cdot {\bf k})^2}.
}
\eqn\JJe
$$
Because of the axial symmetry of the system,
it is reasonable to take ${\bf n}//{\bf r}$,
\foot{
In terms of the energy minimum condition in Eq.{\JJe},
one gets ${\bf n}//{\bf r}$.
Otherwise, there appears the infrared divergence,
corresponding to the double pole in $V_{\rm linear}$.
}
and therefore one finds
$$
\eqalign{
V_{\rm linear}(r)
&=\vec Q^2
\int_{-\infty }^\infty  {dk_r \over 2\pi}\int {d^2k_{_T} \over (2\pi)^2}
\{1-\cos (k_rr)\}{m_B^2 \over k_r^2+k_{_T}^2+m_B^2}
\cdot {1 \over k_r^2}\cr
&={\vec Q^2 m_B^2 \over 8\pi ^2}
\int_{-\infty }^\infty {dk_r \over k_r^2}\{1-\cos(k_rr)\}
\int_0^\infty  dk_{_T}^2 {1 \over k_r^2+k_{_T}^2+m_B^2},
}
\eqn\POTa
$$
where $k_{_T}$ denotes the momentum component perpendicular
to ${\bf r}$.

It should be noted that there appears a physical ultraviolet cutoff
in the $k_{_T}$-integral corresponding to Eq.{\VORTc},
and therefore no ultraviolet divergence comes in Eq.{\POTa}.
The appearance of the physical ultraviolet cutoff is similar
to the argument of the vortex in the superconductivity
[\lifshitz] as mentioned in the previous chapter.
In the central region of the hadron flux tube, $\rho  \lsim m_\chi ^{-1}$,
one find  $m_B \simeq 0$ because the QCD-monopole field
$|\chi |$ almost vanishes. Thus, one gets Eq.{\VORTc},
and there appears the ultraviolet cutoff, $k_{_T} \lsim m_\chi $,
in the integral in Eq.{\POTa} similar to the argument
of the vortex in the superconductivity [\lifshitz].

Hence, we obtain the linear potential,
$$
\eqalign{
V_{\rm linear}(r)
&={\vec Q^2 m_B^2 \over 8\pi ^2}\int_{-\infty }^\infty
{dk_r \over k_r^2}\{1-\cos(k_rr)\}
\ln({m_\chi ^2+k_r^2+m_B^2 \over k_r^2+m_B^2})\cr
&={\rm Re}{\vec Q^2 m_B^2 \over 8\pi ^2}\int_{-\infty }^\infty
{dk_r \over k_r^2}(1-e^{ik_rr})
\ln({m_\chi ^2+k_r^2+m_B^2 \over k_r^2+m_B^2})\cr
&={\rm Re}{\vec Q^2 m_B^2 \over 8\pi ^2}
\pi i{1-e^{ik_rr} \over k_r}
\ln({m_\chi ^2+k_r^2+m_B^2 \over k_r^2+m_B^2})|_{k_r=0}\cr
&={\vec Q^2 m_B^2 \over 8\pi }\ln({m_B^2+m_\chi ^2 \over m_B^2}) \cdot r,
}
\eqn\POTb
$$
where we have used the orthodox technique of the complex integration.
The final expression of the quark static potential is given by
$$
\eqalign{
V(r)&=-{\vec Q^2 \over 4\pi }{e^{-m_Br}\over r}
+{\vec Q^2 m_B^2 \over 8\pi }\ln({m_B^2+m_\chi ^2 \over m_B^2})\cdot r,
}
\eqn\POTc
$$
where $\vec Q^2$ takes the same value for the red, blue and
green quarks, $\vec Q^2=Q_3^2+Q_8^2={e^2 \over 3}$ [\kerson].
We obtain a simple formula for the string tension,
$$
k={\vec Q^2 m_B^2 \over 8\pi }\ln({m_B^2+m_\chi ^2 \over m_B^2}),
\eqn\STR
$$
which is analogous to the energy per unit length of the vortex
in a type-II superconductor,
$$
\epsilon =({\phi _0 \over 4\pi \delta })^2 \ln(\delta /\xi )
={\phi _0^2 m_A^2\over 32\pi ^2}\ln({m_\phi ^2 \over m_A^2}),
\eqn\VOR
$$
where $\phi _0$ is a magnetic flux of the vortex,
$\delta =m_A^{-1}$ and $\xi =m_\phi ^{-1}$ denote the penetration depth
and the coherent length, respectively [\nambuB, \lifshitz].

It is worth mentioning that our expression for the quark confining
potential Eq.{\POTc} differs from the result of the Kanazawa group
[\suzuki,\maedanA].
Their analytic expression for the quark potential includes three
parts, and looks rather complicated.
(See Eq.(4.9) in Ref.{\maedanA}.)
Moreover, their expression for the string tension
is given by the modified Bessel function, $K_0(x)$, and
differs from our expression, Eq.{\STR}.
Such difference stems on the usage of the following
``prescription" by the Kanazawa group,
which used an artificial replacement as
$$
{1 \over (n \cdot k)^2} \rightarrow
{1 \over 2}[{1 \over (n \cdot k+ia)^2}+{1 \over (n \cdot k-ia)^2}]
\eqn\KANAZAWAa
$$
to avoid the infrared double pole in the gluon propagator
\REF\kamizawa{
S.~Kamizawa, Doctor thesis, Kanazawa Univ., (1993).
}
[\maedanA,\kamizawa], where $a$ cannot be taken to the zero limit.
However, this replacement seems troublesome and gives several
difficulties as will be shown in chapter 6. (See Eq.(6.1).)
Moreover, there does {\it not} appear the infrared double pole in the
expression of the quark potential in Eq.{\POTa},
owing to the axial symmetry of the system, ${\bf n}//{\bf r}$.
Hence, there is {\it no need} to eliminate the ``double pole" in this
subject, which is similar to the vortex in the superconductivity:
there does not appear the unphysical divergence related to the
infrared double pole in the vortex solution.
The replacement {\KANAZAWAa} of the Kanazawa group
is unnecessary to obtain the quark potential.

We compare the static quark potential, Eq.{\POTc}, with the
phenomenological potential, for instance,
the Cornell potential
\REF\lucha{
For a recent review article, \nextline
W.~Lucha, F.~F.~Sch\"oberl and D.~Gromes,
Phys.~Rep.~{\bf 200}, No.4 (1991) 127.
}
[\lucha]
in Fig.1.
%
%
The quark potential is thus reproduced in our theory
by choosing  $e=5.5$ (the gauge coupling constant),
$v=126$MeV and $\lambda $=25 in Lagrangian {\ZWANa},
which provide $g=2.3$,
$m_B=500{\rm MeV}$ (mass of the dual gauge field $\vec B_\mu $)
and $m_\chi =1.26 {\rm GeV}$ (QCD-monopole mass).
This parameter set gives $k = 1.0 {\rm GeV/fm}$ for the
string tension in Eq.{\STR}, and the radius of the hadron flux tube
$m_B^{-1} \simeq 0.4 {\rm fm}$.
The linear part of the quark potential is responsible for
the quark confinement and characterizes
the nonperturbative feature in the infrared region.

We discuss now the nonperturbative effect on the quark potential
in the relatively short distance,
$r \sim 0.2{\rm fm} \simeq (1{\rm GeV})^{-1}$, by comparing the
results of the dual Ginzburg-Landau theory and the Cornell potential.
The Cornell potential has the Coulomb part and the linear part,
$$
V_{\rm Cornell}(r)=-{e_{_C}^2 \over 3\pi }\cdot {1 \over r}+k_{_C} r,
\eqn\CORNELL
$$
where the parameters, $e_{_C} \simeq 2$ and
$k_{_C} \simeq 1{\rm GeV/fm}$, are chosen to reproduce the lattice
QCD data in the pure gauge case
and several data of the heavy quarkonium [\lucha].
In the Cornell potential, $k_{_C}$ in the linear part is nothing
but the string tension $k$.
On the other hand, the Coulomb part is introduced to represent
the perturbative one-gluon exchange effect, and therefore
$e_{_C}$ seems to correspond to the gauge coupling constant.
Surely, the Coulomb potential obtained from the
one-gluon exchange would be valid for quite a short distance
as $r \ll 0.2 {\rm fm}$.
In the region of $r \sim 0.2 {\rm fm} \simeq (1{\rm GeV})^{-1}$,
there remains the strong coupling, $\alpha _S \sim 0.5$,
so that the perturbative theory should not be workable
and there must be some contribution beyond the one-gluon
exchange effect.
In this respect, $e_{_C}$ does not correspond to
the gauge coupling constant of QCD directly, but
should be regarded as a fitting parameter
including the higher-order effect of the perturbation or the
nonperturbative effect.

There should be a nonperturbative effect even in the region of
$r \sim 0.2 {\rm fm} \simeq (1{\rm GeV})^{-1}$ in the dual
Ginzburg-Landau theory, because of the presence of
QCD-monopole condensation, whose energy scale is characterized by $m_\chi $.
This nonperturbative effect is found as a reduction of the effective
degrees of freedom of gluons due to QCD-monopole condensation,
because the off-diagonal gluons are considered to lose their
activities in this infrared region.
Although there are $N_c^2-1$ gluons in the SU($N_c$) nonabelian
gauge theory, only $N_c-1$ diagonal gluon fields are considered
to have a large contribution to the infrared region in the dual
Ginzburg-Landau theory.
Hence, the color charge seems to be reduced
to the effective charge in the one-gluon exchange form,
$$
e_{\rm eff}^2
={\sum_{\beta  \in {\rm diag}}
(Q_\beta ^2) e^2 \over \sum_\alpha  (Q_\alpha ^2)}
={(N_c-1) e^2 \over N_c^2-1}={e^2 \over N_c+1},
\eqn\EFFa
$$
where $\sum_{\beta  \in {\rm diag}}$ denotes the summation over
the diagonal component alone.
Then, the parameter $e_{_C}$ in the Coulomb part of the Cornell
potential in Eq.{\CORNELL} should correspond to
$e_{\rm eff}={e \over 2}$
in the dual Ginzburg-Landau theory with the $N_c$=3 case.
This relation is also obtained by the formal
comparison between the Yukawa part in Eq.{\POTc}
and the Coulomb part in Eq.{\CORNELL}
at a short distance.

The derivation of the quark potential {\POTc} is based on
the quenched approximation, where we have neglected
the dynamics of the quark field.
However, in the real QCD system, effects of the dynamical quark
should be important in the infrared (long-range) region of the
potential, because the dynamical $q$-$\bar q$ pair creation should
take place as the hadronic string becomes longer than a critical value.
Due to the $q$-$\bar q$ pair-creation effect, the linear potential
between quark and antiquark is screened in the infrared
(long-distance) region, and the quark static potential
should be asymptotically saturated.
Such an effect has been observed in the lattice QCD simulation
with dynamical quarks [\unger].
In the next chapter, we study the
$q$-$\bar q$ pair creation phenomena and its effect on the
quark confining potential.

\chapter{The $q$-$\bar q$ Pair Creation,
Infrared Screening Effect on the Quark Confinement Potential
and Infrared Cutoff in the Gluon Propagator}

In this chapter, we study the dynamical effect of the light quarks,
particularly the infrared screening on the quark potential due to the
dynamical $q$-$\bar q$ pair creation in the hadron flux tube.
In the real strong interaction system, there are two or three
kinds of light quarks (u,d and s-quark).
The introduction of such light dynamical quarks is important
not only for the dynamical chiral-symmetry breaking [\cheng],
but also for the color confinement, because
the quark confining potential is screened in the infrared
region [\unger] due to the $q$-$\bar q$ pair creation [\suganumaA].

In order to obtain the $q$-$\bar q$ pair creation rate,
we introduce the dynamical quark field interacting with
the color-electric field, which is approximated as a
homogeneous external field inside the hadron flux tube.
The Lagrangian including the quark and the external abelian gluons,
$\vec A^\mu =(A^\mu _3, A^\mu _8)$, is given by
$$
{\cal L}=\bar q (i\not \partial-e\not A-M)q,
\eqn\LAGPa
$$
where $M$ is the effective quark mass,
and $A_\mu \equiv \vec A_\mu  \cdot \vec H$.
By integrating out the quark field in the partition functional,
we obtain the effective action
\REF\itzkson{
For instance,
C.~Itzykson and J.~B.~Zuber, ``Quantum Field Theory'',
(McGraw-Hill, New York, 1985) 1.
}
[\itzkson],
$$
\eqalign{
S_{\rm eff}
&=-{i \over 2}{\rm Tr}\ln\{-(i\partial_\mu -eA_\mu )^2
-{e \over 2}\sigma _{\mu \nu }f^{\mu \nu }+M^2-i\epsilon \} \cr
&={i \over 2} \int d^4x \int_0^\infty {d\tau  \over \tau }{\rm tr}
\langle x|\exp[-i\tau \{-(i\partial_\mu -eA_\mu )^2
-{e \over 2}\sigma _{\mu \nu }f^{\mu \nu }+M^2-i\epsilon \}]|x \rangle
}
\eqn\EFFPa
$$
with $\sigma _{\mu \nu }\equiv{1 \over 2i}[\gamma _\mu , \gamma _\nu ]$ and
$f_{\mu \nu }\equiv \partial_\mu  A_\nu -\partial_\nu  A_\mu $.
After some calculation [\suganumaA],
the effective Lagrangian is given by
$$
{\cal L}_{\rm eff}
={i N_f \over 2\pi }\int_0^\infty  {d\tau  \over \tau }
e^{-i\tau (M^2-i\epsilon )}\int {d^2p_{_T} \over (2\pi )^2}e^{-i\tau p_{_T}^2}
{\rm tr}_c(eE)\coth(eE\tau ).
\eqn\EFFPb
$$
Here, $E$ is a  {\rm diagonal matrix} with the color indices
corresponding to the magnitude of the color-electric field,
$E=\{(E_3 T_3 +E_8 T_8 )^2\}^{1/2}$, where
$E_3=(E_3^iE_3^i)^{1/2}$ and $E_8=(E_8^iE_8^i)^{1/2}$ are
the magnitude of the two kinds of color-electric field,
$\vec E^i=(E_3^i, E_8^i)=\partial^0 \vec A^i-\partial^i \vec A^0
(i=1,2,3)$.
The integration variable $p_{_T}$ physically means the
momentum component of dynamical quarks
perpendicular to the direction of the
color-electric field.
By using the Wick rotation in the complex $\tau $-plane,
one obtains the Schwinger formula for the $q$-$\bar q$ pair
creation rate $w$
[\suganumaA],
$$
\eqalign{
w&=2 {\rm Im}{\cal L}_{\rm eff}
  ={N_f \over \pi }\sum_{l=1}^\infty
   \int{d^2p_{_T} \over (2\pi )^2} {1 \over l}
   {\rm tr}_c(eE)e^{-l\pi (p_{_T}^2+M^2)(eE)^{-1}} \cr
&=-{N_f \over 2\pi ^2}
   \int_0^\infty  dp_{_T} p_{_T}
   {\rm tr}_c(eE)\ln\{1-e^{-\pi (p_{_T}^2+M^2)(eE)^{-1}}\} \cr
&={N_f \over 4\pi ^3}\sum_{l=1}^\infty
   {1 \over l^2}{\rm tr}_c(eE)^2e^{-l\pi M^2(eE)^{-1}},
}
\eqn\PAIRa
$$
which is purely a nonperturbative effect in terms of the
gauge coupling constant $e$, and cannot be obtained by
the use of the perturbation with respect to $e$.

Let us consider the expression of $eE$ in the hadron flux tube with
the cross section $S$, $S \sim \pi  m_B^{-2} \sim 0.5 {\rm fm}^2$.
In a fundamental representation of SU(3)$_c$,
the red ($R$), blue ($B$) and green ($G$) quarks are
expressed by (1,0,0), (0,1,0), (0,0,1), respectively.
In this representation, the color charge $\vec Q $ is
given by
$
(Q_3, Q_8)=( {1 \over 2}e, {1 \over 2\sqrt{3} }e),
           (-{1 \over 2}e, {1 \over 2\sqrt{3} }e),
           (            0, -{1 \over  \sqrt{3} }e)
$
for the red, blue and green quark, respectively
[\kerson].
Then, the diagonal matrix $eE$, which corresponds to the
magnitude of the color-electric field, is given by
$$
eE=e\{ (E_3 T_3  +  E_8 T_8)^2 \}^{1/2}
  ={e \over S} \{( Q_3 T_3 + Q_8 T_8)^2\}^{1/2}
  ={\rm diag}(2k,k,k)
\eqn\Ea
$$
in the flux tube of the $R$-$\bar R$ system.
Here, we have used the relation on the string tension,
$k={\vec Q^2 \over 2S}={e^2 \over 6S} (\simeq 1 {\rm GeV/fm})$,
which is derived from the Gauss law $\vec Q=\vec E S$, and
$
k={1 \over 2}\vec E^2 S ={1 \over 2}(E_3^2+E_8^2)S.
$
\foot{
Here, the spatial variation of the QCD-monopole field
is neglected because it only appears in the central region
of the flux tube, $\rho  \lsim m_\chi ^{-1}$.
}
Similarly, one finds $eE={\rm diag}(k,2k,k), {\rm diag}(k,k,2k)$
for the $B$-$\bar B$ and $G$-$\bar G$ system, respectively.

Hence, one finds the same expression for $w$
in the $R$-$\bar R$, $B$-$\bar B$ and $G$-$\bar G$ system,
$$
\eqalign{
w&=-{N_f \over 2\pi ^2}
   \int_0^\infty  dp_{_T} p_{_T}
   [2k \ln\{1-e^{-\pi (p_{_T}^2+M^2)/(2k)}\}
   +2 \cdot k \ln\{1-e^{-\pi (p_{_T}^2+M^2)/k}\}] \cr
&={N_f \over 4\pi ^3}\sum_{l=1}^\infty
   {1 \over l^2}
   [(2k)^2e^{-l\pi M^2/(2k)}+2 \cdot k^2e^{-l\pi M^2/k}].
}
\eqn\PAIRb
$$
The first term in the bracket corresponds to the $q$-$\bar q$
pair creation with the same color as the valence quark,
and such a pair-creation process contributes to the cut of the hadron
flux tube, {\it i.e.} the screening effect of the linear potential
between the valence quarks.
As Glendenning and Matsui
\REF\glendenning{
N.~K.~Glendenning and T.~Matsui,
Phys.~Rev.~{\bf D28} (1983) 2890; \nextline
Phys.~Lett.~{\bf B141} (1984) 419.
}[\glendenning]
pointed out,
the effective color charge of the created $q$-$\bar q$ pair
becomes a half value due to the screening effect or the final-state
interaction in this process, and $2k$ in the first term of
Eq.{\PAIRb} is reduced to $k$.
On the other hand, the second term in Eq.{\PAIRb} describes the case
that the color of the created $q$-$\bar q$ pair differs from that of the
valence quark, so that this contribution is
less important to the screening of the quark potential.
Thus, the $q$-$\bar q$ pair creation rate relevant
to the screening effect is given by
$$
w_{\rm sc}=-{N_f \over 2\pi ^2}k
   \int_0^\infty  dp_{_T} p_{_T}
     \ln\{1-e^{-\pi (p_{_T}^2+M^2)/k}\}
 ={N_f \over 4\pi ^3}k^2
   \sum_{l=1}^\infty  {1 \over l^2}e^{-l\pi M^2/k}.
\eqn\PAIRc
$$

We estimate now the expectation value of the energy of the
created $q$-$\bar q$ pair system.
The creation rate of the $q$-$\bar q$ pair with
the transverse momentum $p_{_T}$ is given by
$$
w(p_{_T})=-{N_f \over 2\pi ^2}k
      p_{_T} \ln\{1-e^{-\pi (p_{_T}^2+M^2)/k}\},
\eqn\PAIRd
$$
and therefore the expectation value of the created-pair energy
is estimated by
$$
\langle 2 E_q \rangle={1 \over w_{\rm sc}}
          \int_0^\infty  dp_{_T} w(p_{_T}) \cdot 2 E_q(p_{_T}),
\eqn\EVEV
$$
where $E_q(p_{_T})$ denotes the energy of a quark with the
transverse momentum $p_{_T}$ and is estimated as
$E_q(p_{_T}) \simeq (p_{_T}^2+M^2)^{1/2}$.
\foot{
Since $p_{_T}^2=p_x^2+p_y^2$,
it may be reasonable to use
$E_q(p_{_T}) \simeq (p^2+M^2)^{1/2}
=(p_x^2+p_y^2+p_z^2+M^2)^{1/2}
\simeq ({3 \over 2} p_{_T}^2+M^2)^{1/2}$.
However, the peak or the average of the $q$-$\bar q$ pair
creation rate $w(p_{_T})$ is modified only slightly in this case.
}

We show in Fig.2 the $q$-$\bar q$ pair creation rate
vs. the created-pair energy $2 E_q$.
Here, we have used $k = 1.0 {\rm GeV/fm}$ and
$M = 350{\rm MeV}$ for the string tension and the effective
mass of the quark, respectively.
\foot{
The quark effective mass $M \simeq 350{\rm MeV}$ is obtained
in the low-momentum region
by using the Schwinger-Dyson equation in our framework
as will be shown in the next chapter.
}
We obtain the expectation value of the created pair energy,
$\langle 2 E_q \rangle \simeq $ 850MeV.
Since the created pair energy $\langle 2 E_q \rangle $ is supplied
by the shortening of the hadronic string by cutting off the string,
the screening effect due to the $q$-$\bar q$ pair creation
appears for the longer distance than $R_{\rm sc}$ such that
$k R_{\rm sc} \simeq \langle 2 E_q \rangle $.
Then, we obtain the screening distance $R_{\rm sc} \simeq 1{\rm fm}$.

The hadronic strings become unstable against
the $q$-$\bar q$ pair creation when the distance between the valence
quark and antiquark becomes larger than $R_{\rm sc}$.
This means the vanishing of the strong correlation between the quark
and the antiquark in the infrared region
due to the screening effect brought by the light $q$-$\bar q$ pair
creation and the cut of the hadronic string.
Then, the corresponding infrared cutoff,
$\epsilon  \simeq R_{\rm sc}^{-1} \simeq 200{\rm MeV}$,
should be introduced to the long-range strong correlation between the
valence quarks.

In the dual Ginzburg-Landau theory, the long range
correlation between quarks is brought by the nonlocal operator
${1 \over (n \cdot \partial)^2}$
in the diagonal-gluon propagator in Eq.{\LPd}.
{}From the one-dimensional relation
$
\partial{1 \over \partial}(x)
=\partial\theta (x)=\delta (x),
$
one finds that
$
{1 \over \partial}(x)=\theta (x)+C_1
$
with a constant $C_1$, and similarly one obtains
$$
{1 \over \partial^2}(x)=x\theta (x)+C_1x+C_2,
\eqn\TETAc
$$
where $C_1$ and  $C_2$ are arbitrary constants.
Thus, one finds that
$$
\eqalign{
{1 \over (n \cdot \partial)^2} (x-y)
&\equiv \langle x | {1 \over (n \cdot \partial)^2} | y \rangle \cr
&=\{(x_n-y_n)\theta (x_n-y_n)+C_1(x_n-y_n)+C_2\}
\delta ^3(x_{_T}-y_{_T}),
}
\eqn\SLRC
$$
which gives a larger correlation for a further point
to the direction $n_\mu $.
Such an operator brings the strong long-range correlation
along $n_\mu $, direction of the Dirac string,
and leads to the linear confining potential
in the absence of the dynamical quarks.

In the presence of light dynamical quarks,
the infrared cutoff $\epsilon  \simeq 200{\rm MeV}$ should be introduced to
this long-range correlation factor,
${1 \over (n \cdot \partial)^2}$, in the gluon propagator,
in accordance with the $q$-$\bar q$ pair creation and the
cut of the hadronic string.
\foot{
In principle, the infrared screening effect in the gluon propagator
would be calculable by solving the Schwinger-Dyson
equation including the polarization effect of dynamical quarks.
}
We get a modified diagonal-gluon propagator
including the infrared screening effect as
$$
D_{\mu \nu }^{\rm sc}(k)=-{1 \over k^2}\{g_{\mu \nu }+(\alpha _e-1)
{k_\mu  k_\nu  \over k^2} \}
+{1 \over k^2}\cdot
{m_B^2 \over k^2-m_B^2}\cdot
{n^2 \over (n\cdot k)^2+\epsilon ^2}X_{\mu \nu },
\eqn\MGO
$$
in the momentum representation from Eq.{\LPd}.
\foot{
Although there is another way to introduce the infrared
cutoff $\epsilon $, they qualitatively give the same results on the
infrared screening effect in the quark potential.
}
Hence, the corresponding inverse propagator is given by
$$
D_{\mu \nu }^{\rm sc} \ ^{-1}(k)=
-\{g_{\mu \nu }k^2-(1-{1 \over \alpha _e}) k_\mu  k_\nu  \}
-{k^2 m_B^2 n^2 \over
 k^2\{(n\cdot k)^2-m_B^2n^2\}+\epsilon ^2(k^2-m_B^2)}X_{\mu \nu },
\eqn\MGOI
$$
and the modified Lagrangian reads
$$
{\cal L}_{\rm DGL-MF}^{\rm sc}
  = {1 \over 2} \vec A^\mu  D_{\mu \nu }^{\rm sc} \ ^{-1} \vec A^\nu
   +\bar q (i\not \partial-e \not \vec A \cdot \vec H-m)q,
\eqn\EFFLagz
$$
which holds the [U(1)]$_e^2$-gauge invariance when the gauge fixing
term is absent, {\it i.e.} $\alpha _e=\infty $.
It is notable that the infrared double pole, e.g.
${1 \over (n\cdot k)^2}$, naturally disappears
in the gluon propagator $D_{\mu \nu }^{\rm sc}(k)$,
which comes to play an important role to the study of the dynamical
chiral-symmetry breaking as will be shown in the chapter 6.
Then, the quark potential $V_{\rm linear}(r)$
in Eq.{\POTb} is modified by
$$
\eqalign{
V_{\rm linear}^{\rm sc}(r)
&\simeq {\vec Q^2 m_B^2 \over 8\pi ^2}\int_{-\infty }^\infty
{dk_r \over (k_r^2+\epsilon ^2)}\{1-\cos(k_rr)\}
\ln({m_\chi ^2+k_r^2+m_B^2 \over k_r^2+m_B^2})\cr
&={\rm Re}{\vec Q^2 m_B^2 \over 8\pi ^2}\int_{-\infty }^\infty
{dk_r \over (k_r^2+\epsilon ^2)}(1-e^{ik_rr})
\ln({m_\chi ^2+k_r^2+m_B^2 \over k_r^2+m_B^2})\cr
&={\rm Re}{\vec Q^2 m_B^2 \over 8\pi ^2}
2\pi i {1-e^{ik_rr} \over k_r+i\epsilon }
\ln({m_\chi ^2+k_r^2+m_B^2 \over k_r^2+m_B^2})|_{k_r=i\epsilon }\cr
&={\vec Q^2 m_B^2 \over 8\pi }\cdot
{1-e^{-\epsilon r} \over \epsilon }
\ln({m_B^2+m_\chi ^2 -\epsilon ^2\over m_B^2-\epsilon ^2})
}
\eqn\POTSCa
$$
apart from a small contribution of $O(\epsilon ^2)$.

We obtain a compact formula for the quark
potential including the infrared screening effect due to the
$q$-$\bar q$ pair creation,
$$
V_{\rm sc}(r)=V_{\rm Yukawa}(r)+V_{\rm linear}^{\rm sc}(r)
=-{\vec Q^2 \over 4\pi }\cdot {e^{-m_Br}\over r}
+k \cdot {1-e^{-\epsilon r} \over \epsilon },
\eqn\POTc
$$
where $k$ corresponds to the string tension,
$$
k={\vec Q^2 \over 8\pi }[(m_B^2-\epsilon ^2)
\ln({m_B^2+m_\chi ^2 -\epsilon ^2\over m_B^2-\epsilon ^2})
+\epsilon ^2 \ln({m_\chi ^2-\epsilon ^2 \over \epsilon ^2})].
\eqn\STRSC
$$
The quark potential $V_{\rm linear}^{\rm sc}(r)$
behaves as the linear potential with the string tension $k$,
$$
V_{\rm linear}^{\rm sc}(r) \simeq k r
\eqn\POTSCb
$$
at the short distance, $r \ll \epsilon ^{-1}$.
On the contrary, the quark potential shows a saturation
corresponding to the screening effect,
$$
V_{\rm linear}^{\rm sc}(r)
\simeq {k \over \epsilon }={\rm constant}
\eqn\POTSCb
$$
in the infrared region, $r \gg \epsilon ^{-1}$.

We show the numerical result of
$V_{\rm linear}^{\rm sc}(r)$ for several values
of infrared cutoff $\epsilon $ in Fig.3
as the screening effect on the quark confining potential.
We have used the same values for the gauge coupling constant
and the mass of the dual gauge field as before,
$e=5.5$ and $m_B =500{\rm MeV}$.
The QCD-monopole mass $m_\chi $ is chosen so as to reproduce the
string tension $k = 1.0 {\rm GeV/fm}$, and is found to be
$m_\chi $ = 1.26, 1.15 and 1.01 GeV for $\epsilon $=0, 100 and 200 MeV,
respectively.
The screening effect or the saturation of the quark
confining potential is found in the long distance, and
such a screening effect becomes dominant as $\epsilon $ gets larger.

Let us consider the relationship among the quantities
related to the quark confinement and the infrared screening effect.
As for the infrared cutoff $\epsilon $ and the
effective quark mass $M$, we find a simple relation
$$
\epsilon  \simeq {k \over \langle 2E_q \rangle} \sim {k \over 2M}
\eqn\CUTa
$$
from the relations, $\epsilon  \simeq R_{\rm sc}^{-1}$ and
$kR_{\rm sc} \simeq \langle 2E_q \rangle \sim 2M$.
We also find a simple relation between the saturated value
of the confining potential, $V_{\rm linear}^{\rm sc}(\infty )$, and
the string tension $k$,
$$
V_{\rm linear}^{\rm sc}(\infty )={k \over \epsilon }
\simeq  \langle 2E_q \rangle \sim 2M
\eqn\RELAa
$$
from Eqs.{\STRSC}, {\POTSCb} and {\CUTa}.

We examine the dependence of the effective quark mass $M$
within the possible variation of the effective quark mass,
$350{\rm MeV} \lsim M< \infty $,
because even the massless quark has a large effective mass
about $350 {\rm MeV}$ due to the dynamical chiral-symmetry breaking.
{}From Eqs.{\STRSC} and {\CUTa},
one finds that $\epsilon $ is almost inversely proportional to $M$,
and the string tension $k$ does not strongly depend on $\epsilon $
and $M$ as long as $\epsilon  \ll m_B, m_\chi $.
One also finds that the saturated value
$V_{\rm linear}^{\rm sc}(\infty )$ is almost proportional to the
quark mass $M$ from Eq.{\RELAa}.
These relations can be examined by the use of the lattice QCD
simulation for the various bare quark mass.

The infrared screening effect on the quark potential has been
observed in the lattice QCD simulation with the dynamical quarks
[\unger].
However, most studies have been based on the small-size lattice,
for instance $8^3 \times 4$, and the finite size effect
may spoil the quantitative argument on the long-range screening
effect. Actually, rather different results are reported
on the screening effect in the lattice QCD data [\unger].
It is desirable to investigate this problem by using a
large-size lattice with high statistics.

\chapter{The Schwinger-Dyson Equation and
the Dynamical Chiral-Symmetry Breaking}

In this chapter, we study the dynamical chiral-symmetry breaking
by using the dual Ginzburg-Landau theory.
In particular, we are interested in the relation between
the color confinement and the dynamical chiral-symmetry breaking
because there is a remarkable coincidence on the critical temperatures
of the deconfinement phase transition and the chiral-symmetry
restoration in the lattice QCD simulation with dynamical quarks
\REF\petersson{
For instance,
B.~Petersson, Nucl.~Phys.~{\bf A525} (1991) 273c;
published in Amsterdam Lattice 1992 (1992) 66
and references therein.
}
[\petersson].
To this end, we derive the Schwinger-Dyson equation for
the light quarks by using the diagonal-gluon propagator
$D_{\mu \nu }^{\rm sc}(k)$ in Eq.{\MGO} in the dual Ginzburg-Landau theory.

Several years ago, Baker et al.
\REF\bakerA{
M.~Baker, J.~S.~Ball and F.~Zachariasen,
Phys.~Rev.~{\bf D38} (1988) 1926;
Erratum-{\it ibid.} {\bf D47} (1993) 743.
}
\REF\kreinA{
G.~Krein and  A.~G.~Williams,
Phys.~Rev.~{\bf D43} (1991) 3541.
}
[\bakerA, \kreinA]
studied the relation between the dynamical chiral-symmetry
breaking and monopole condensation in the dual QCD model
\REF\bakerB{
For a recent review article, \nextline
M.~Baker, J.~S.~Ball and F.~Zachariasen,
Phys.~Rep.~{\bf 209} (1991) 73.
}
[\bakerB],
and obtained that monopole condensation contributes
to the chiral symmetry breaking.
However, they used a {\it wrong sign} [\bakerA, \kreinA] for the
integrand in the Schwinger-Dyson equation, which was pointed out
by the Kanazawa group [\kamizawa].
Since this error was fatal as was shown in erratum in Ref.[\bakerA],
their conclusion is no more creditable.
Very recently, the Kanazawa group studied this subject by using
the dual Ginzburg-Landau theory, and
they encountered a difficulty in solving the Schwinger-Dyson
equation in relation with the appearance of the infrared
double pole in the gluon propagator [\kamizawa].
They used an artificial replacement {\KANAZAWAa}
to avoid the infrared double pole [\maedanA,\kamizawa],
and used an angle average for the direction of the Dirac string $n_\mu $
[\kamizawa,\kreinA],
$$
{1 \over 2\pi ^2} \int d\Omega _n {1 \over (n \cdot k)^2}=-{2 \over k^2}.
\eqn\ANGLEav
$$
However, the inverse of the sign takes place in this replacement
because the integrand in the left-hand side is positive definite.
They solved the Schwinger-Dyson equation by using the above
replacement, and found that the dynamical chiral-symmetry breaking
is weakened by QCD-monopole condensation [\kamizawa].
Their result seems unexpected and against the results on the
relationship between the deconfinement phase transition and
the chiral-symmetry restoration in the lattice QCD [\petersson].
We speculate that the inverse of the sign
in the replacement {\ANGLEav} is unnatural and problematic,
and gives the origin of their unnatural result.

We shall now study this issue by using the gluon propagator
including the infrared screening effect
in the dual Ginzburg-Landau theory.
Although the Kanazawa group has neglected the dynamical effect of light
quarks, it is desirable, as an important ingredient,
to include the dynamical effect of quarks such as the
$q$-$\bar q$ pair creation in this subject,
because the light dynamical quarks should be introduced
for the study of the dynamical chiral-symmetry breaking.
Hence, we use the modified diagonal-gluon propagator
$D_{\mu \nu }^{\rm sc}(k)$ in Eq.{\MGO}, which
includes the infrared screening effect.
It is notable that the infrared double pole in the
gluon propagator is naturally disappeared by the infrared
screening effect. (See Eq.{\MGO}.)
Hence, there is no difficulty of the infrared double pole
in the gluon propagator, and we can formulate the Schwinger-Dyson
equation in a straightforward way.

We study the dynamical chiral-symmetry breaking
by using the Schwinger-Dyson equation for the quark field
in the chiral limit within the rainbow approximation
[\higashijima, \miransky],
$$
S^{-1}_q(p_{_M})=\not p_{_M}+ \int {d^4k_{_M} \over i(2\pi )^4}
\vec Q^2 \gamma ^\mu S_q(k_{_M})\gamma ^\nu D_{\mu \nu }^{\rm
sc}(k_{_M}-p_{_M}),
\eqn\SDEa
$$
where $S_q(p_{_M})$ denotes the quark propagator with
the Minkowski momentum $p_{_M}$.
In this equation, the coupling strength
between the quark and the diagonal-gluons is given by
$\vec Q^2={N_c-1 \over 2N_c}e^2$
in the SU($N_c$) gauge theory [\kamizawa].
We consider a simple quark propagator $S_q(p)$ including the
momentum-dependent mass $M(-p_{_M}^2)$,
$$
S_q(p_{_M})^{-1}=\not p_{_M} -M(-p_{_M}^2)+i\epsilon ,
\eqn\QUARKa
$$
although the quark propagator should be taken, in a rigorous sense, as
\nextline
$
S_q(p_{_M})^{-1}=A \not p_{_M} +B +C
\not p_{_M} \not n+D\not n +i\epsilon ,
$
where $A, B, C,$ and $D$ are scalar functions of $p_{_M}^2$ and
$p_{_M} \cdot n$ [\kreinA].
The Schwinger-Dyson equation for the dynamical quark mass
$M(-p_{_M}^2)$ is obtained by taking the trace of Eq.{\SDEa},
$$
M(p^2)= \int{d^4k \over (2\pi )^4} \vec Q^2 {M(k^2) \over k^2+M^2(k^2)}
       D_\mu ^{\mu  {\rm sc}}(k-p)
\eqn\SDEb
$$
after the Wick rotation in the $k_0$-plane
and the transformation into the Euclidean variables,
which are simply denoted as $p$ or $k$.
In the Lorentz gauge, one finds
$$
\eqalign{
D_\mu ^{\mu  {\rm sc}}(k)&=
     {1 \over (n \cdot k)^2+\epsilon ^2} \cdot {1 \over k^2} \cdot
     {2m_B^2 \over k^2+m_B^2}\{k^2-(n \cdot k)^2 \}
     +{3+\alpha _e \over k^2} \cr
&={2 \over (n \cdot k)^2+\epsilon ^2}
\left(
{m_B^2 \over k^2+m_B^2}+{\epsilon ^2 \over k^2}-{\epsilon ^2 \over k^2+m_B^2}
\right)
+{2 \over k^2+m_B^2}+{1+\alpha _e \over k^2},
}\eqn\MGOa
$$
and the Schwinger-Dyson equation becomes
$$
\eqalign{
M(p^2)
&= \int{d^4k \over (2\pi )^4} \vec Q^2 {M(k^2) \over k^2+M^2(k^2)} \cr
&\times
\biggl(
{2 \over \tilde k^2+m_B^2}+{1+\alpha _e \over \tilde k^2}
+{2 \over (n \cdot \tilde k)^2+\epsilon ^2}
[{m_B^2 \over \tilde k^2+m_B^2}+{\epsilon ^2 \over \tilde k^2}
-{\epsilon ^2 \over \tilde k^2+m_B^2} ]
\biggr)
}
\eqn\SDEc
$$
with $\tilde k_\mu  \equiv k_\mu -p_\mu $.
One finds that no infrared double pole appears in Eq.{\SDEc},
so that one can solve it numerically without any difficulties related
to the unphysical infrared divergence.

We can perform the angle integration of the
first and second terms in the bracket in Eq.{\SDEc}
by the use of the formula,
$$
\eqalign{
I_1
&\equiv
\int{d^4k \over (2\pi )^4}{1 \over \tilde k^2+c^2}f(k^2)\cr
&={1 \over (2\pi )^3}\int_0^\infty k^2dk^2 f(k^2)
\int_0^\pi  d\theta {\sin^2 \theta  \over k^2+p^2+c^2-2kp\cos\theta }\cr
&={1 \over 8\pi ^2}\int_0^\infty k^2dk^2 f(k^2)
{1 \over k^2+p^2+c^2+\sqrt{(k^2+p^2+c^2)^2-4k^2p^2}},
}
\eqn\INTEGRALa
$$
where $f(k^2)$ and $c$ are an arbitrary function of $k^2$
and an arbitrary constant, respectively.
As for the remaining part in Eq.{\SDEc},
we can partially perform the angle integration using the
formula,
$$
\eqalign{
I_2
&\equiv
\int{d^4k \over (2\pi )^4} {1 \over (n \cdot \tilde k)^2+\epsilon ^2}
\cdot {1 \over \tilde k^2+c^2}f(k^2) \cr
&=\int_{-\infty }^\infty {dk_n \over 2\pi }
  \int_0^\infty  {k_{_T}^2dk_{_T} \over (2\pi )^2} f(k^2)
  {1 \over \tilde k_n^2+\epsilon ^2}
  \int_{-1}^1
  {d(\cos\alpha ) \over
  \tilde k_n^2+k_{_T}^2+p_{_T}^2+c^2-2k_{_T}p_{_T}\cos\alpha } \cr
&={1 \over 16\pi ^3 p_{_T}}
  \int_{-\infty }^\infty  dk_n
  \int_0^\infty  dk_{_T} k_{_T} f(k^2)
  {1 \over \tilde k_n^2+\epsilon ^2}
  \ln \left({\tilde k_n^2+(k_{_T}+p_{_T})^2+c^2 \over
  \tilde k_n^2+(k_{_T}-p_{_T})^2+c^2 }\right) \cr
&={1 \over 32\pi ^3 p_{_T}}
  \int_0^\infty  dk^2 f(k^2)
  \int_{-k}^k dk_n
  {1 \over \tilde k_n^2+\epsilon ^2}
  \ln \left({\tilde k_n^2+(k_{_T}+p_{_T})^2+c^2 \over
  \tilde k_n^2+(k_{_T}-p_{_T})^2+c^2 } \right),
}
\eqn\INTEGRALb
$$
where we have used the relation,
$$
 \int_{-\infty }^\infty dk_n\int_0^\infty  dk_{_T}k_{_T}
=\int_{-\infty }^\infty dk_n\int_{|k_n|}^\infty  dk k
=\int_0^\infty dk k \int_{-k}^k dk_n.
\eqn\INTEGRALc
$$

Then, after some complicated calculation on the angle integrations,
one obtains a expression for the Schwinger-Dyson equation,
$$
\eqalign{
M(p^2)
 &=\int_0^\infty  {dk^2 \over 16\pi ^2} {\vec Q^2 M(k^2) \over k^2+M^2(k^2)}
  \biggl(
 {4k^2   \over   k^2+p^2+m_B^2 + \sqrt{(k^2+p^2+m_B^2)^2-4k^2p^2} }
 \cr
 &+{(1+\alpha _e)k^2 \over {\rm max}(k^2, p^2)}
  +{1 \over \pi p_{_T}} \int_{-k}^k dk_n {1 \over \tilde k_n^2+\epsilon ^2} \cr
 &\times [ (m_B^2-\epsilon ^2) \ln \{ { {\tilde k_n^2+(k_{_T}+p_{_T})^2+m_B^2
   \over \tilde k_n^2+(k_{_T}-p_{_T})^2+m_B^2} } \}
  +\epsilon ^2  \ln \{ { {\tilde k_n^2+(k_{_T}+p_{_T})^2
   \over \tilde k_n^2+(k_{_T}-p_{_T})^2} } \}   ] \biggr),
}
\eqn\SDEd
$$
where $\tilde k_n \equiv k_n-p_n$, $k_{_T}\equiv(k^2-k_n^2)^{1/2}$.
Here, the right hand side of Eq.{\SDEd} is a function of
$p_n$ and $p_{_T}$ generally,
so that Eq.{\SDEd} is not a closed equation,
which is brought due to the $n_\mu $-dependence of the diagonal
gluon propagator in Eq.{\MGO}.
To solve Eq.{\SDEd}, we use the ansatz,
$$
p_n=p \sin \theta , \quad  p_{_T}= p \cos \theta
\eqn\ANSATZa
$$
with an angle parameter $\theta $,
and then the Schwinger-Dyson equation {\SDEd}
can be solved for each value of $\theta $.
It is notable that the right hand side of Eq.{\SDEd} is
always {\it non-negative}, which supports the existence of a
nontrivial solution.

In solving the Schwinger-Dyson equation,
we use the Higashijima-Miransky approximation
[\higashijima, \miransky]
with a hybrid type of the running coupling constant
\REF\atkinson{
D.~Atkinson and P.~W.~Johnson,
Phys.~Rev.~{\bf D37} (1988) 2296.
}
[\kamizawa,\atkinson]
$$
\tilde e=e({\rm max}\{p^2, k^2\}), \quad
e^2(p^2)=
{48\pi ^2 (N_c+1) \over (11N_c-2N_f)
\ln\{(p^2+p^2_c)/\Lambda ^2_{\rm QCD}\}},
\eqn\GCa
$$
and replace the gauge coupling constant $e$ with $\tilde e$
in Eq.{\SDEd}.
Here, $\Lambda _{\rm QCD }$ is the QCD scale parameter
($\Lambda _{\rm QCD}(\overline{\rm MS})=220 \pm 15 \pm 50{\rm MeV}$, Exp.)
[\bcdms] and $p_c$ is given by
$$
p_c^2=\Lambda _{\rm QCD}^2
\exp[{48\pi ^2 \over e^2 } \cdot {N_c+1 \over (11N_c-2N_f)}].
\eqn\ICa
$$
Such a treatment of the gauge coupling constant,
$\tilde e$ in Eq.{\GCa}, approximately divides
the momentum region into the ultraviolet part
$p>p_c$ and the infrared part $p<p_c$.
In the high momentum region, $p^2 \gg p_c^2$,
gauge coupling strength in Eq.{\SDEd} is reduced to
the perturbative one in Eq.{\RCCa},
$$
{\vec Q^2(p^2) \over 4\pi }={N_c-1 \over 2N_c} \cdot {e^2(p^2) \over 4\pi }
\simeq {N_c^2-1 \over 2N_c}
{12\pi   \over (11N_c-2N_f) \ln(p^2/\Lambda ^2_{\rm QCD})}
={N_c^2-1 \over 2N_c}\alpha _s(p^2),
\eqn\HCb
$$
and therefore Eq.{\SDEd} is reduced to a simple expression
in the ultraviolet limit $p^2\rightarrow \infty $,
$$
\eqalign{
M(p^2)
= {N_c^2-1 \over 2N_c}
  \int_0^\infty  {dk^2 \over 4\pi }
 \alpha _s ({\rm max}\{k^2, p^2\}){M(k^2) \over k^2+M^2(k^2)} \cdot
 {(3+\alpha _e)k^2 \over {\rm max}(k^2, p^2)},
}
\eqn\SDEe
$$
which coincides with the Schwinger-Dyson equation
by using the free gluon field [\higashijima, \miransky].
Thus, the full gluon contribution is taken into account
in this ultraviolet region by the use of $\tilde e$
in Eq.{\GCa} [\kamizawa].
On the other hand, $e(p^2)$ in Eq.{\GCa} becomes a constant
in the low momentum region, $p^2  \lsim p_c^2$,
$$
e^2(p^2) \simeq {48\pi ^2 (N_c+1) \over (11N_c-2N_f) \ln(p_c^2/\Lambda _{\rm
QCD}^2 )}
         =e^2,
\eqn\GCb
$$
which seems underestimation of the strong-coupling nature
in comparison with the perturbative result in Eq.{\RCCa}.
It should be noted that the dual Ginzburg-Landau theory
includes QCD-monopole condensation as a
{\it nonperturbative effect }
related to the color confinement, which would be brought
by the strong coupling in the infrared region.

We briefly show our thinking on the nonperturbative QCD
physics in terms of the dual Ginzburg-Landau theory.
Owing to the asymptotic freedom of QCD, the perturbation theory
is valid in the ultraviolet region, while the dynamics in the
infrared region should be nonperturbative.
In the dual Ginzburg-Landau theory, these two regions
are approximately divided by the energy scale of the dual Meissner
effect, the mass of the dual-gauge field $m_B$ or the abelian
monopole mass $m_\chi $, and therefore $p_c$ in the above argument
should correspond to $m_B (m_\chi )$, {\it i.e.} $p_c \sim m_B (m_\chi )$,
which is valid for our parameter set,
$e=5.5, \Lambda _{\rm QCD}$=200MeV and $N_c=N_f=3$,
$p_c \simeq 638{\rm MeV} \sim m_B (m_\chi )$.
Thus, in our framework, the ultraviolet region of
$p>p_c \sim m_B (m_\chi )$ is regarded as the perturbative region
and is characterized by the running coupling constant
in the perturbation theory.
On the other hand, the infrared region of $p<p_c \sim m_B (m_\chi )$
is essentially nonperturbative and is characterized by abelian
monopole condensation or the dual Meissner effect
in the dual Ginzburg-Landau theory.

We study the dynamical chiral-symmetry breaking by solving the
Schwinger-Dyson equation {\SDEd} numerically in the
Landau gauge $\alpha _e=0$, where the wave function renormalization
is not needed [\kugo] and the quark propagator {\QUARKa} is valid
for $m_B=0$.
As for the angle parameter $\theta $, we obtain similar numerical solutions
for various $\theta $ and the $\theta $-dependence seems rather small,
and therefore we only show the numerical results for the averaged
values of $p_n$ and $p_{_T}$,
$p_n=\langle p_n \rangle={1 \over 2}p$ and
$p_{_T}=\langle p_{_T} \rangle={\sqrt{3} \over 2}p$.
We show the numerical result for the dynamical quark mass $M(p^2)$
in Fig.4 for $e=5.5$, $\Lambda _{\rm QCD}=200{\rm MeV}$,
$m_B=500{\rm MeV}$ and $\epsilon $=80MeV.
The solid curve in the space-like region, $p^2(=-p_{_M}^2)>0$,
is a nontrivial solution of the Schwinger-Dyson equation {\SDEd},
where the dynamical quark mass in the infrared region,
$M(0) \simeq 348{\rm MeV}$, seems a reasonable value
in terms of the constituent quark model [\nachtmann].

We also investigate the light-quark confinement,
although this issue is rather difficult because
the discussion of the {\rm static} quark potential cannot be
applied to the light quarks unlike the heavy quarks
because the light quark cannot be fixed at a spatial point.
Instead, the light-quark confinement is characterized by the
disappearance of the physical poles in the quark propagator,
which can be obtained by the {\it analytic continuation} of
the dynamical quark mass $M(p^2)$
to the time-like region in principle
\REF\fukuda{
For the QED version, \nextline
R.~Fukuda and T.~Kugo, Nucl.~Phys.~{\bf B117} (1976) 250.
}
[\fukuda].
In Fig.4, the solid curve in the time-like region,
$p^2(=-p_{_M}^2)<0$, is obtained by a smooth extrapolation
of $M(p^2)$ from the region $0<p^2<p_c^2$,
assuming the continuous property of the quark propagator.
We find the solid curve does not cross the line,
$p^2(=-p_{_M}^2)=-M^2(p^2)$, corresponding to the on-shell condition.
Hence, the quark propagator has no physical poles,
and the light quarks never appear as the asymptotic fields,
which means the color confinement for {\it light quarks}.

We now investigate the effect of QCD-monopole condensation
on the dynamical chiral-symmetry breaking.
We show in Fig.5 the numerical results of the dynamical
quark mass $M(p^2)$ for the various values of $m_B$,
the mass of the dual-gauge field $\vec B_\mu $.
The other parameters are set to the same values as in Fig.4.
No nontrivial solution is found for the small values
of $m_B$, e.g.  $m_B \lsim 200 {\rm MeV}$.
The nontrivial solution is found for $m_B \gsim 300{\rm MeV}$,
and the dynamical quark mass $M(p^2)$ becomes larger
at each $p^2$ as $m_B$ gets larger, and therefore
QCD-monopole condensation provides a large contribution to the
dynamical chiral-symmetry breaking.
Thus, we find the close relation between the color confinement
and the dynamical chiral-symmetry breaking through QCD-monopole
condensation in the dual Ginzburg-Landau theory.

We also calculate the quark condensate $\langle \bar qq \rangle$,
and the pion decay constant $f_\pi $, which also
characterize the dynamical chiral-symmetry breaking
as well as the dynamical quark mass in the infrared limit, $M(0)$.
The quark condensate is given by
$$
\langle \bar qq \rangle_\Lambda
=\int^\Lambda  {d^4k_{_M} \over i(2\pi )^4}{\rm tr} S_q(k_{_M})
=-{N_c \over 4\pi ^2}\int_0^{\Lambda ^2} dk^2 {k^2 M(k^2) \over k^2+M^2(k^2)},
\eqn\QUCONa
$$
where ultraviolet cutoff $\Lambda $ is introduced to regularize
the ultraviolet divergence of the integral.
By using the asymptotic form of the quark mass $M(p^2)$
\REF\lane{
K.~Lane, Phys.~Rev.~{\bf D10} (1974) 2605. \nextline
H.~D.~Politzer, Nucl.~Phys.~{\bf B117} (1976) 397. \nextline
H.~Pagels, Phys.~Rev.~{\bf D19} (1979) 3080.
}
[\lane],
one gets the quark condensate in the renormalization-group
invariant form in the asymptotic region
\REF\kreinB{
G.~Krein, P.~Tang and A.~G.~Williams,
Phys.~Lett.~{\bf B215} (1988) 145.
}
[\kamizawa,\kreinB],
$$
\langle \bar qq \rangle_{_{\rm RGI}}=
{\langle \bar qq \rangle_\Lambda   \over \{\ln(\Lambda ^2/\Lambda ^2_{\rm
QCD})\}^a },
\eqn\RGIa
$$
where $a={9(N_c^2-1) \over 2N_c (11N_c-2N_f)}
={4 \over 9}$ for $N_c=N_f=3$.
We calculate $\langle \bar qq \rangle_{_{\rm RGI}}$
in the asymptotic region $\Lambda ^2/\Lambda _{\rm QCD}^2=10^6$,
where the electro-weak unification takes place.
The pion decay constant ($f_\pi $=93MeV, Exp.) is given by
the Pagels-Stokar formula,
\REF\pagels{
H.~Pagels and S.~Stoker, Phys.~Rev.~{\bf D20} (1979) 2947;
{\bf D22} (1980) 2876.
}
[\pagels]
$$
f_\pi ^2={N_c \over 4\pi ^2}\int_0^\infty dk^2
{k^2 M(k^2) \over \{k^2+M^2(k^2)\}^2}
\left(
M(k^2)-{k^2 \over 2}\cdot {dM(k^2) \over dk^2}
\right).
\eqn\FPIa
$$

The numerical results on $M(0)$,
$\langle \bar qq \rangle_{_{\rm RGI}}$
and $f_\pi $ are shown in Fig.6 for the various values of
the gauge coupling $e$ and the mass of the dual-gauge field $m_B$.
As for the $m_B$ dependence on these quantities,
the dynamical-symmetry breaking appears strongly for the large value
of $m_B$, and therefore a large amount of the chiral-symmetry
breaking seems to be brought by QCD-monopole condensation,
which is responsible to the color confinement.
These quantities are well reproduced as
$M(0) \simeq 348{\rm MeV}$,
$\langle \bar qq \rangle_{_{\rm RGI}} \simeq-(192{\rm MeV})^3$,
$\langle \bar qq \rangle_{\Lambda =1{\rm GeV}}\simeq-(229{\rm MeV})^3$
and $f_\pi  \simeq 83.6{\rm MeV}$
by using the parameter set,
$e=5.5$, $\Lambda _{\rm QCD}$=200 MeV [\bcdms], $m_B$=500MeV and $\epsilon
$=80MeV.
This parameter set is consistent with the perturbative QCD
in the ultraviolet region [\nachtmann] and the quark potential
as shown in chapter 4 and 5.
In particular, this parameter set provides the reasonable values
of the running gauge coupling constant,
$\alpha _s(m_Z^2) \simeq 0.123 ({\rm Exp.}, 0.108 \pm 0.005 )$ [\amaldi]
and $\alpha _s((35{\rm GeV})^2)\simeq 0.146 ({\rm Exp.}, 0.10 \sim 0.17)$
[\nachtmann] for $N_f=4$.
Thus, the dual Ginzburg-Landau theory provides
the consistent picture for the phenomenological aspects of QCD,
that is the perturbative QCD, the color confinement and
the dynamical chiral-symmetry breaking.

It is notable that
large values for $e$ and $\Lambda _{\rm QCD}$ are needed
in the absence of QCD-monopole condensation, $m_B=0$,
to reproduce the enough amount of the chiral-symmetry breaking,
as is shown in Fig.6.
This case, $m_B=0$, corresponds to the Schwinger-Dyson equation
in Refs.[\barducci] and [\kugo], where the authors used
large values for
$e \sim 10$ and $\Lambda _{\rm QCD}=0.5 \sim 1 {\rm GeV}$
to reproduce $\langle \bar qq \rangle_{_{\rm RGI}}$ or $f_\pi $.
This seems inconsistent with the QCD scale parameter
$\Lambda _{\rm QCD} \sim 200{\rm MeV}$ obtained from the high-energy
experimental data by using the perturbative QCD [\bcdms],
and the gauge coupling $e$ estimated from the static quark potential.
(See chapter 4.)
The reason of the disagreement for $m_B=0$ seems
rather trivial, because there is no nonperturbative effect
like the color confinement in the infrared region,
where the coupling constant $e$ becomes extremely
strong and the nonperturbative effect should appear.
Such a missing  contribution in the infrared region
inevitably leads the underestimation for
the physical quantities like $\langle \bar qq \rangle$ or $f_\pi $.
Hence, if one wants to reproduce enough values for them for $m_B=0$,
one needs a large coupling $e \sim 10$ and a
large QCD scale parameter $\Lambda _{\rm QCD}= 0.5 \sim 1{\rm GeV}$,
which enlarges the strong-coupling region ($p<\Lambda _{\rm QCD}$).

Finally, we consider the physical meaning of the above results
on the close relation between the confinement and the
dynamical chiral-symmetry breaking through
QCD-monopole condensation.
The dynamical chiral-symmetry breaking is characterized by
the quark and antiquark pair condensation,
$\langle \bar qq \rangle \ne 0$, which is brought by the
strong attractive force between the quark and the antiquark
with the opposite color charge [\shuryak, \nambuA].
In the dual Ginzburg-Landau theory, QCD-monopole condensation
leads the linear confining potential, which bring
the strong and long-range attractive force between
the quark and the antiquark with the opposite color charge,
and therefore quark pair condensation is expected to be realized.
Thus, the dynamical chiral-symmetry breaking should be
enhanced by the presence of the confining force
or QCD-monopole condensation.
On the other hand, only a Coulomb-type force remains
between quarks when the QCD-monopole field is not condensed,
$m_B=0$, and therefore the weak attractive force
is only expected between the quark and the antiquark.
Thus, the realization of the dynamical chiral-symmetry breaking
is rather hard in the absence of QCD-monopole condensation,
unless quite large values for $e$ and $\Lambda _{\rm QCD}$ are used,
as shown in Fig.6.

\chapter{Summary and Discussions}

We have studied nonperturbative features in QCD such as
the color confinement, the $q$-$\bar q$ pair creation
and the dynamical chiral-symmetry breaking
in terms of QCD-monopole condensation
by using the dual Ginzburg-Landau theory.
We have investigated the appearance of QCD-monopoles in QCD
in the abelian gauge proposed by 't~Hooft.
The nonabelian nature leads to the nontrivial homotopy class,
$\pi _2({\rm SU}(N_c)/{\rm [U(1)]}^{N_c-1})
=\pi _1({\rm [U(1)]}^{N_c-1})=Z_\infty ^{N_c-1}$,
which provides the origin of the QCD-monopoles.
As a phenomenological theory of the nonperturbative QCD,
we have constructed the dual Ginzburg-Landau theory
including the confining mechanism in terms of the
dual Meissner effect by QCD-monopole condensation.
As for the heavy quark confinement,
we have derived the static quark potential including
the linear part and the Yukawa part
in the dual Ginzburg-Landau theory
within the quenched approximation.
We have derived a simple formula for the string tension,
which is analogous to the energy per unit length of the vortex
in the superconductivity.

We have studied the dynamical effect of light quarks on the quark
potential with respect to the infrared screening effect brought by
the $q$-$\bar q$ pair creation or the cut of the hadronic strings.
We have estimated the screening length, $R_{sc} \simeq 1{\rm fm}$,
for the quark confining potential by the use of the Schwinger
formula for the $q$-$\bar q$ pair creation.
The corresponding infrared cutoff has been introduced to the
strong long-range correlation part in the gluon propagator,
and we have obtained a compact formula for the quark potential
including the screening effect in the infrared region.

We have investigated the dynamical chiral-symmetry breaking
in the dual Ginzburg-Landau theory
by the use of the Schwinger-Dyson equation with the gluon
propagator including the dual Meissner effects or
the nonperturbative effect related to the color confinement.
In our approach,
the nonperturbative region and the perturbative region
are naturally divided by the energy scale of QCD-monopole
condensation, and characterized by the dual Meissner effect and
the perturbative running coupling constant, respectively.
We have found a large enhancement of the dynamical chiral-symmetry
breaking due to QCD-monopole condensation, which supports
the close relation between the color confinement and the
chiral-symmetry breaking through QCD-monopole condensation.

The dynamical quark mass, the pion decay constant and the quark
condensate are well reproduced by using the consistent values
of the gauge coupling constant and the QCD scale parameter with
the perturbative QCD and the quark confining potential.
By using the  parameter set, $e = 5.5$, $\Lambda _{\rm QCD} =
200{\rm MeV}$, $m_B =500{\rm MeV}$ and $\epsilon $=80MeV,
we have obtained good results on both the quark confining potential
and the dynamical chiral-symmetry breaking, e.g.
$M(0) \simeq 348{\rm MeV}$,
$\langle \bar qq\rangle_{_{\rm RGI}} \simeq -(192{\rm MeV})^3$,
$\langle \bar qq\rangle_{\Lambda =1{\rm GeV}} \simeq -(229{\rm MeV})^3$,
and $f_\pi  \simeq 83.6{\rm MeV}$.
This parameter set also provides the reasonable values
of the running gauge coupling constant,
$\alpha _s(m_Z^2) \simeq 0.123 ({\rm Exp.}, 0.108 \pm 0.005 )$
and $\alpha _s((35{\rm GeV})^2)\simeq 0.146 ({\rm Exp.}, 0.10 \sim 0.17)$
for $N_f=4$, and is found to be consistent with the perturbative QCD
in the ultraviolet region.

The color confinement of the light quarks has been also examined
in the dual Ginzburg-Landau theory
by the smooth extrapolation of the quark mass function $M(p^2)$
from the space-like region to the time-like region, and
the disappearance of the physical poles in the light-quark propagator
has been found.

Thus, we have obtained the {\it consistent picture }
for the various phenomena in QCD by using the dual
Ginzburg-Landau theory. We have shown in this theory
the quark confining potential within the quenched approximation,
the infrared screening effect for the quark potential by the
light quark-pair creation,
the dynamical chiral-symmetry breaking by using
the consistent value of $\Lambda _{\rm QCD}$ with the perturbative QCD,
and the light-quark confinement.

We now discuss the other interesting subjects in QCD in terms of
QCD-monopole condensation in the dual Ginzburg-Landau theory.
It is meaningful to examine the relation between abelian
monopole condensation and the QCD phase transition,
which is characterized by the confinement and the chiral symmetry.
Due to the asymptotic freedom of QCD,
the gauge coupling constant becomes small at high temperatures,
and one expects the phase transition from the nonperturbative vacuum
to the perturbative vacuum, which corresponds to the deconfinement
phase transition and the chiral-symmetry restoration [\shuryak].
The critical temperatures of these phase transitions
coincide in the study of the
lattice QCD simulation with light dynamical quarks [\petersson],
and the critical gauge couplings also
coincide between the deconfinement and the chiral-symmetry
restoration in the lattice QCD.
Hence, there must be close relation between the confinement and
the chiral-symmetry breaking.

In the dual Ginzburg-Landau theory,
the QCD phase transition is characterized by the presence or
the absence of QCD-monopole condensation.
In the strong-coupling case or the low-temperature case,
we expect QCD-monopole condensation, which results in
the color confinement and the dynamical chiral-symmetry breaking.
At a critical coupling or a critical temperature,
such QCD-monopole condensation would disappear,
and therefore one expects not only the disappearance of the
color confinement, e.g. the dual Meissner effect or the
confining force between quarks, but also the large reduction
of the dynamical chiral-symmetry breaking.
In particular, the dynamical chiral-symmetry breaking also
vanishes for $e \lsim 7$ when the QCD-monopole condensate
disappears, $m_B=0$. (see Fig.6.)
Thus, we expect the coincidence of the deconfinement phase
transition and the chiral-symmetry restoration
in relation with the vanishing of QCD-monopole condensation.

We will briefly show the future program of our study in the
dual Ginzburg-Landau theory.
It is straightforward to extend the Schwinger-Dyson equation
for the massive current quark, $m \ne 0$, and therefore
we can also investigate the properties related to
the massive quarks such as s, c and  t-quark, e.g. s-quark mass
$M_s(p^2)$, $f_K$ and $\langle \bar ss \rangle$,
as well as u and d-quark [\barducci] in the dual Ginzburg-Landau
theory.

The hadron properties can be studied by solving
the Bethe-Salpeter equation
[\kugo]
or the Faddeev equation
\REF\belyaev{
For a recent review article,
V.~B.~Belyaev, ``Lectures on the Theory of Few-Body Systems",
(Springer-Verlag, Berlin ,1990) 1.
}
[\belyaev] in the dual Ginzburg-Landau theory.
The mass and the decay constant of mesons ($\sigma $, $\rho $, $A_1$)
would be calculated by using
the similar technique of the ladder Bethe-Salpeter equation
in Ref.[\kugo] with the diagonal-gluon propagator
$D_{\mu \nu }^{\rm sc}(k)$ in Eq.{\MGO},
although the equation may be complicated due to its $n_\mu $ dependence.
It would be interesting to investigate the properties of
mesons including the heavy quark by using the Bethe-Salpeter
equation in this scheme, because dynamics of the c-quark is hardy
controlled in the lattice gauge simulation.
Although the baryon property can be obtained by the Faddeev equation
in principle, it would be rather complicated to solve numerically.
However, approximate properties of the baryon would be calculable
by regarding the baryon as the bound state of the quark and
the diquark
\REF\anselmino{
Papers in Proc. of the Workshop on ``Diquarks",
ed. M.~Anselmino et al.
(World Scientific, Singapore, 1988) 1.
}
\REF\stech{
B.~Stech, Phys.~Rev.~{\bf D36} (1987) 975;
Nucl.~Phys.~{\bf B}(Proc.~Suppl.){\bf 7a} (1989) 106.
}
[\anselmino, \stech].

We consider the other possible applications
of the dual Ginzburg-Landau theory to the hadron physics.
The nonrelativistic quark models
\REF\mukherjee{
For a recent review article,
S.~N.~Mukherjee, R.~Nag, S.~Sanyal, T.~Morii, J.~Morishita
and M.~Tsuge, Phys.~Rep.~{\bf 231} (1993) 201.
}
[\mukherjee]
may be reformulated by using the diagonal-gluon propagator
in the dual Ginzburg-Landau theory,
instead of the introduction of the confining potential,
which would be automatically reproduced.
The chiral soliton picture [\skyrme] of the nucleon or the
$\Delta $(1232)-resonance based on the nontrivial homotopy group
$\pi _2({\rm SU}(N_f)/{\rm U}(1))=\pi _1({\rm U}(1))=Z_\infty $
is one of the most interesting issue in the hadron physics,
and such a chiral soliton would be investigated
from the underlying quark degrees of freedom
by using the $n_\mu $-averaged quark-current correlation
in Eq.{\LPb} and the similar argument to the soliton
in the Nambu-Jona-Lasinio model
\REF\goeke{
T.~Meissner, F.~Grummer and K.~Goeke, Phys.~Lett.~{\bf B227} (1989) 296.
}
[\goeke].

We have divided into the nonperturbative and the
perturbative regions in terms of QCD-monopole condensation
in the dual Ginzburg-Landau theory.
We compare such a division with the operator product expansion
\REF\wilson{
K.~G.~Wilson, Phys.~Rev.~{\bf 179} (1969) 1499.
}[\wilson], where
one introduces normalization point $\mu $,
and fluctuations with virtual low momentum $p$ such as
$p^2<\mu ^2$ are included in the local operators, while
those at large virtual momentum $p^2>\mu ^2$ are included
in the Wilson coefficients by definition [\shuryak].
Since the small scale physics is approximately described
by the perturbation theory and nonperturbative phenomena are
connected with the long scale physics,
the normalization point $\mu $ physically corresponds to
the border of the nonperturbative and the perturbative regions.
In our framework, such a division is given by the typical
energy scale of QCD-monopole condensation,
$p_c \sim m_B (m_\chi ) \sim 1{\rm GeV}$, and
the running coupling constant based on a perturbative
scheme is used for the small scale physics, $p^2>p_c^2$,
while the nonperturbative nature appears as
QCD-monopole condensation mainly
in the long scale physics, $p^2<p_c^2$.
Thus, our framework seems similar to the operator product expansion,
and then QCD-monopole condensation
may provide the physical image of the separation between the
two different scale regions in the operator product expansion.

Furthermore, the dual Ginzburg-Landau theory would also
provide the {\it calculation scheme} for
the matrix elements of the local operators,
e.g. $\langle G_{\mu \nu }G^{\mu \nu }\rangle $ or $\langle \bar qq \rangle $,
in the operator product expansion, because we have obtained
the quark propagator and the diagonal-gluon propagator
including the nonperturbative effect.
Hence, a powerful technique should be obtained
by using such a calculation for the matrix elements of the
local operators and the operator product expansion
in the QCD sum rule
\REF\shifman{
M.~A.~Shifman, A.~I.~Vainshtein
and V.~I.~Zakharov, Nucl.~Phys. {\bf B147} (1979) 385, 448 and 519.
\nextline
L.~J.~Reinders, H.~Rubinstein and S.~Yazaki,
Phys.~Rep.~{\bf 127} (1985) 1 and references therein.
}[\shifman]
in estimating the various physical quantities.
Such a calculation scheme would be also applied to the
estimation of the structure functions [\nachtmann]
of the nucleon observed in deep inelastic lepton-nucleon scattering
as a useful technique for the studies of
the proton spin problem
\REF\jaffe{
R.~L.~Jaffe and A.~Manohar, Nucl.~Phys.~{\bf B337} (1990) 509.
}[\jaffe]
and the breaking of the Gottfried sum rule
\REF\nmc{
New Muon Collaboration, Phys.~Rev.~Lett.~{\bf 66} (1991) 2717
and references therein.
}
[\nmc].

\vskip1cm

\centerline{\fourteenpoint ACKNOWLEDGEMENTS}

\vskip0.5cm

We are grateful to H.~Monden for her contributions on QCD-monopole
condensation in the early stage of this study.
We have used the program D7NLFRHM of TLIB in Tokyo Computational
Center made by  K.-I.~Kondo, H.~Nakatani and  H.~Mino
in solving the Schwinger-Dyson equation.
One of the authors (H.S.) is supported by the Special Researchers'
Basic Science Program at RIKEN.

\vskip1cm

\refout

\figout

\end